\documentclass[apjl]{emulateapj}
\usepackage{comment}
\usepackage{ifthen}
\usepackage{lineno}

\usepackage{amsmath}

\newcommand{\ctbd}[1]{}

\newcommand{\lc}{light curve}

\newcommand{\Lc}{Light curve}

\newcommand{\kms}{\ensuremath{\rm km\,s^{-1}}}
\newcommand{\ms}{\ensuremath{\rm m\,s^{-1}}}

\newcommand{\gcmc}{\ensuremath{\rm g\,cm^{-3}}}
\newcommand{\ergscmsq}{\ensuremath{\rm erg\,s^{-1}\,cm^{-2}}}

\newcommand{\teff}{\ensuremath{T_{\rm eff}}}
\newcommand{\logg}{\ensuremath{\log{g}}}
\newcommand{\vsini}{\ensuremath{v \sin{I_\star}}}
\newcommand{\feh}{\ensuremath{\rm [Fe/H]}}

\newcommand{\rsun}{\ensuremath{R_\sun}}
\newcommand{\msun}{\ensuremath{M_\sun}}
\newcommand{\lsun}{\ensuremath{L_\sun}}

\newcommand{\rstar}{\ensuremath{R_\star}}
\newcommand{\mstar}{\ensuremath{M_\star}}
\newcommand{\lstar}{\ensuremath{L_\star}}

\newcommand{\teffstar}{\ensuremath{T_{\rm eff\star}}}

\newcommand{\loggstar}{\ensuremath{\log{g_{\star}}}}

\newcommand{\rpl}{\ensuremath{R_{p}}}
\newcommand{\mpl}{\ensuremath{M_{p}}}

\newcommand{\rhopl}{\ensuremath{\rho_{p}}}

\newcommand{\arstar}{\ensuremath{a/\rstar}}

\newcommand{\rjup}{\ensuremath{R_{\rm J}}}
\newcommand{\mjup}{\ensuremath{M_{\rm J}}}

\newcommand{\reftabl}[1]{Table~\ref{tab:#1}}

\newcommand{\hatcurCCtwomass}{J17062656+4446371}                  
\newcommand{\hatcurCCgsc}{03084-00533}                              
\newcommand{\hatcurCCapassmV}{\ensuremath{10.069\pm0.016}}             
\newcommand{\hatcurLCdur}{\ensuremath{0.364_{-0.091}^{+0.122}}}                
\newcommand{\hatcurLCP}{\ensuremath{4.8101050\pm0.00000044}}              
\newcommand{\hatcurLCPprec}{\ensuremath{4.8101050}}                    
\newcommand{\hatcurLCT}{\ensuremath{2455961.38464\pm0.00099}}          
\newcommand{\hatcurSMEiteff}{\ensuremath{6166\pm50}}                   
\newcommand{\hatcurSMEizfeh}{\ensuremath{-0.2\pm0.1}}                 
\newcommand{\hatcurSMEizfehshort}{\ensuremath{-0.2}}                   
\newcommand{\hatcurSMEilogg}{\ensuremath{4.51\pm0.10}}                 
\newcommand{\hatcurSMEivsin}{\ensuremath{0.8\pm0.5}}                   
\newcommand{\hatcurSMEivmac}{\ensuremath{NULL}}                        
\newcommand{\hatcurSMEivmic}{\ensuremath{NULL}}                        
\newcommand{\hatcurSMEiiteff}{\ensuremath{NULL\pmNULL}}                
\newcommand{\hatcurSMEiizfeh}{\ensuremath{NULL\pmNULL}}                
\newcommand{\hatcurSMEiizfehshort}{\ensuremath{NULL}}                  
\newcommand{\hatcurSMEiilogg}{\ensuremath{NULL\pmNULL}}                
\newcommand{\hatcurSMEiivsin}{\ensuremath{NULL\pmNULL}}                
\newcommand{\hatcurSMEiivmac}{\ensuremath{NULL}}                       
\newcommand{\hatcurSMEiivmic}{\ensuremath{NULL}}                       
\newcommand{\hatcurISOm}{\ensuremath{1.45_{-0.16}^{+0.21}}}                     
\newcommand{\hatcurISOr}{\ensuremath{2.43_{-0.66}^{+0.94}}}                     
\newcommand{\hatcurISOage}{\ensuremath{3.2_{-1.5}^{+0.9}}}             
\newcommand{\hatcurISOspec}{G}                                         
\newcommand{\hatcurRVK}{\ensuremath{38_{-24}^{+30}}}                       
\newcommand{\hatcurPPg}{\ensuremath{2.15_{-0.45}^{+0.36}}}                        
\newcommand{\hatcurPPrho}{\ensuremath{0.026_{-0.018}^{+0.056}}}                    
\newcommand{\hatcurPPm}{\ensuremath{0.38_{-0.24}^{+0.31}}}                      
\newcommand{\hatcurPPr}{\ensuremath{2.59_{-0.72}^{+0.97}}}                      
\newcommand{\hatcurPPteff}{\ensuremath{2890_{-390}^{+430}}}                     
\newcommand{\hatcurXdist}{\ensuremath{257\pm8}}                        

\newcommand{\hatcurCCapassmVcirc}{\ensuremath{10.069\pm0.016}}         
\newcommand{\hatcurCCapassmBcirc}{\ensuremath{10.682\pm0.010}}         
\newcommand{\hatcurCCtwomassJmagcirc}{\ensuremath{9.145\pm0.021}}     
\newcommand{\hatcurCCtwomassHmagcirc}{\ensuremath{8.961\pm0.019}}     
\newcommand{\hatcurCCtwomassKmagcirc}{\ensuremath{8.900\pm0.019}}     
\newcommand{\hatcur}{HAT-P-67}
\newcommand{\hatcurb}{HAT-P-67b}

\newcommand{\hatcurSMEversion}{i}                                       
\newcommand{\hatcurSMEteff}{\ifthenelse{\equal{\hatcurSMEversion}{i}}{\hatcurSMEiteff}{\hatcurSMEiiteff}}
\newcommand{\hatcurSMEzfeh}{\ifthenelse{\equal{\hatcurSMEversion}{i}}{\hatcurSMEizfeh}{\hatcurSMEiizfeh}}
\newcommand{\hatcurSMEzfehshort}{\ifthenelse{\equal{\hatcurSMEversion}{i}}{\hatcurSMEizfehshort}{\hatcurSMEiizfehshort}}
\newcommand{\hatcurSMElogg}{\ifthenelse{\equal{\hatcurSMEversion}{i}}{\hatcurSMEilogg}{\hatcurSMEiilogg}}
\newcommand{\hatcurSMEvsin}{\ifthenelse{\equal{\hatcurSMEversion}{i}}{\hatcurSMEivsin}{\hatcurSMEiivsin}}
\newcommand{\hatcurSMEvmac}{\ifthenelse{\equal{\hatcurSMEversion}{i}}{\hatcurSMEivmac}{\hatcurSMEiivmac}}
\newcommand{\hatcurSMEvmic}{\ifthenelse{\equal{\hatcurSMEversion}{i}}{\hatcurSMEivmic}{\hatcurSMEiivmic}}

\newcommand{\hatcurSMEversioneccen}{i}                                       
\newcommand{\hatcurSMEteffeccen}{\ifthenelse{\equal{\hatcurSMEversioneccen}{i}}{\hatcurSMEiteffeccen}{\hatcurSMEiiteffeccen}}
\newcommand{\hatcurSMEzfeheccen}{\ifthenelse{\equal{\hatcurSMEversioneccen}{i}}{\hatcurSMEizfeheccen}{\hatcurSMEiizfeheccen}}
\newcommand{\hatcurSMEzfehshorteccen}{\ifthenelse{\equal{\hatcurSMEversioneccen}{i}}{\hatcurSMEizfehshorteccen}{\hatcurSMEiizfehshorteccen}}
\newcommand{\hatcurSMEloggeccen}{\ifthenelse{\equal{\hatcurSMEversioneccen}{i}}{\hatcurSMEiloggeccen}{\hatcurSMEiiloggeccen}}
\newcommand{\hatcurSMEvsineccen}{\ifthenelse{\equal{\hatcurSMEversioneccen}{i}}{\hatcurSMEivsineccen}{\hatcurSMEiivsineccen}}
\newcommand{\hatcurSMEvmaceccen}{\ifthenelse{\equal{\hatcurSMEversioneccen}{i}}{\hatcurSMEivmaceccen}{\hatcurSMEiivmaceccen}}
\newcommand{\hatcurSMEvmiceccen}{\ifthenelse{\equal{\hatcurSMEversioneccen}{i}}{\hatcurSMEivmiceccen}{\hatcurSMEiivmiceccen}}

\newcommand{\hatcurisoshort}{Geneva}
\newcommand{\hatcurisocite}{2012A&A...537A.146E}
\newcommand{\hatcurjhkfilset}{ESO}

\newboolean{emulateapj}
\setboolean{emulateapj}{true}

\newboolean{rvtablelong}
\setboolean{rvtablelong}{true}

\newboolean{astroph}
\setboolean{astroph}{true}

\newcommand{\genevastarmass}{\ensuremath{1.642_{-0.072}^{+0.155}}}                 
\newcommand{\genevastarradius}{\ensuremath{2.546_{-0.084}^{+0.099}}}                 
\newcommand{\genevaplanetmass}{\ensuremath{0.34_{-0.19}^{+0.25}}}                 
\newcommand{\genevaplanetradius}{\ensuremath{2.085_{-0.071}^{+0.096}}}                 
\newcommand{\genevastarlogg}{\ensuremath{3.854_{-0.023}^{+0.014}}}              
\newcommand{\genevastarvsini}{\ensuremath{35.8\pm1.1}}             
\newcommand{\genevastarteff}{\ensuremath{6406_{-61}^{+65}}}              
\newcommand{\genevalambda}{\ensuremath{2.9_{-4.9}^{+6.4}}}              
\newcommand{\genevaplrho}{\ensuremath{0.052_{-0.028}^{+0.039}}}         

\shortauthors{Zhou et al.}
\shorttitle{\hatcur\lowercase{b}}
\ifthenelse{\boolean{emulateapj}}{
    
}{
    
}
\ifthenelse{\boolean{emulateapj}}{
    
}{
    
}
\ifthenelse{\boolean{emulateapj}}{
    
}{
    
}

\begin{document}


\title{\hatcur\lowercase{b}: an extremely low density Saturn transiting an F-subgiant confirmed via Doppler tomography  \altaffilmark{$\dagger$}}

\author{
G. Zhou\altaffilmark{1},
G. \'A. Bakos\altaffilmark{2,$\star$},
J. D. Hartman\altaffilmark{2},
D. W. Latham\altaffilmark{1},
G. Torres\altaffilmark{1},
W. Bhatti\altaffilmark{2},
K. Penev\altaffilmark{2},
L. Buchhave\altaffilmark{3},
G. Kov\'acs\altaffilmark{4},
A. Bieryla\altaffilmark{1},
S. Quinn\altaffilmark{1},
H. Isaacson\altaffilmark{5},
B. J. Fulton\altaffilmark{6},
E. Falco\altaffilmark{1},
Z. Csubry\altaffilmark{2},
M. Everett\altaffilmark{7},
T. Szklenar\altaffilmark{8},
G. Esquerdo\altaffilmark{1},
P. Berlind\altaffilmark{1},
M. L. Calkins\altaffilmark{1},
B.~B\'eky\altaffilmark{9},
R.~P.~Knox\altaffilmark{10},
P.~Hinz\altaffilmark{10},
E.~P.~Horch\altaffilmark{11},
L.~Hirsch\altaffilmark{5},
S.~B.~Howell\altaffilmark{12},
R. W. Noyes\altaffilmark{1},
G. Marcy\altaffilmark{5},
M. de Val-Borro\altaffilmark{2},
J. L\'az\'ar\altaffilmark{8},
I. Papp\altaffilmark{8},
P. S\'ari\altaffilmark{8}
}
\altaffiltext{1}{Harvard-Smithsonian Center for Astrophysics, Cambridge,
MA 02138, USA}
\altaffiltext{2}{Department of Astrophysical Sciences, Princeton
University, Princeton, NJ 08544, USA}
\altaffiltext{$\star$}{Packard Fellow}
\altaffiltext{3}{Centre for Star and Planet Formation, Natural History
Museum of Denmark, University of Copenhagen, DK-1350
Copenhagen, Denmark}
\altaffiltext{4}{Konkoly Observatory of the Hungarian Academy of Sciences,
Budapest, Hungary}
\altaffiltext{5}{Department of Astronomy, University of California, Berkeley,
CA, USA}
\altaffiltext{6}{Institute for Astronomy, University of Hawaii, Honolulu, HI
96822, USA}
\altaffiltext{7}{National Optical Astronomy Observatory, Tucson, AZ, USA}
\altaffiltext{8}{Hungarian Astronomical Association, Budapest, Hungary}
\altaffiltext{9}{Google Inc.}
\altaffiltext{10}{Steward Observatory, University of Arizona, 933 N. Cherry Ave., Tucson, AZ 85721, USA}
\altaffiltext{11}{Department of Physics, Southern Connecticut State University, 501 Crescent Street, New Haven, CT 06515, USA}
\altaffiltext{12}{NASA Ames Research Center, Moffett Field, CA 94035, USA}

\altaffiltext{$\dagger$}{
Based on observations obtained with the Hungarian-made Automated
Telescope Network. Based in part on observations made with the Keck-I
telescope at Mauna Kea Observatory, HI (Keck time awarded through NASA
programs N029Hr, N108Hr, N154Hr and N130Hr and NOAO programs
A289Hr, and A284Hr). Based in part on
observations obtained with the Tillinghast Reflector 1.5 m telescope and the
1.2 m telescope, both operated by the Smithsonian Astrophysical Observatory
at the Fred Lawrence Whipple Observatory in Arizona. This work makes use of the Smithsonian Institution High Performance Cluster (SI/HPC).  Based in part on observations made with the Nordic
Optical Telescope, operated on the island of La Palma jointly by Denmark,
Finland, Iceland, Norway, and Sweden, in the Spanish Observatorio del Roque
de los Muchachos of the Instituto de Astrof\'isica de Canarias.
}


\begin{abstract}

We report the discovery of \hatcurb{}, a hot-Saturn transiting a rapidly rotating F-subgiant. \hatcurb{} has a radius of $R_p = \genevaplanetradius \,R_J$, orbiting a $M_* = \genevastarmass \,M_\odot$, $R_* = \genevastarradius \,R_\odot$ host star in a $\sim 4.81$-day period orbit. We place an upper limit on the mass of the planet via radial velocity measurements to be $M_p < 0.59\,M_J$, and lower limit of $> 0.056\,M_J$ by limitations on Roche lobe overflow. Despite being a subgiant, the host star still exhibits relatively rapid rotation, with a projected rotational velocity of $\vsini = \genevastarvsini \, \kms$, making it difficult to precisely determine the mass of the planet using radial velocities. We validated \hatcurb{} via two Doppler tomographic detections of the planetary transit, which eliminated potential eclipsing binary blend scenarios. The Doppler tomographic observations also confirmed that \hatcurb{} has an orbit that is aligned to within $12^\circ$, in projection, with the spin of its host star. \hatcurb{} receives strong UV irradiation, and is amongst the one of the lowest density planets known, making it a good candidate for future UV transit observations to search for an extended hydrogen exosphere.

\setcounter{footnote}{0}
\end{abstract}

\keywords{
    planetary systems ---
    stars: individual (\hatcur{}, \hatcurCCgsc{}) 
    techniques: spectroscopic, photometric
}


\section{Introduction}
\label{sec:introduction}

Finding well-characterized planets in a variety of environments is key to understanding the processes that govern planet formation and evolution. Planets orbiting high mass stars are likely born in high mass protoplanetary disks \citep[e.g.][]{Muzerolle:2003,Natta:2006}, environments that may yield higher planet occurrence rates \citep[e.g.][]{Johnson:2010,2010ApJ...709..396B} and higher-mass planets \citep[e.g.][]{2007A&A...472..657L,Jones:2014} than around solar type stars. Planets around early type stars also receive higher incident flux over their lifetimes, which in turn make them anchor-points in the planet mass-radius-equilibrium temperature relationships \citep[e.g.][]{2011ApJ...734..109B,2012A&A...540A..99E}. 

However, only 1\% of known transiting planets orbit stars more massive than $1.5\,\msun$. Early type stars have larger radii, resulting in shallower transit depths for any planets; they are also more likely to have rotationally blended spectral lines due to the lack of magnetic braking over the main-sequence lifetime, making traditional radial-velocity confirmation techniques more difficult. One successful strategy is to conduct radial-velocity surveys of `retired A-stars' -- stars that have evolved off the main-sequence and spun down enough to exhibit sharp spectroscopic lines that enable precise radial-velocity measurements. These surveys have been extremely successful, yielding 122 planetary systems to date\footnote{Choosing host stars with $\logg < 4.0$, from NASA Exoplanet Archive, July 2016} \citep[e.g.][]{Johnson:2007,Wittenmyer:2011,Jones:2014}. Recently, transit surveys have also been successful in discovering planets around high mass stars. These include planets around subgiants and giants whose shallow transits were identified by \emph{Kepler} (e.g. Kepler-56b,c \citealt{2013Sci...342..331H}, Kepler-96b \citealt{2014A&A...562A.109L}, Kepler-432b \citealt{2015ApJ...803...49Q,2015A&A...573L...5C}, KOI-206b, KOI-680b \citealt{2015A&A...575A..71A}, and K2-39b \citealt{2016arXiv160509180V}), and hot-Jupiters around main-sequence A-stars confirmed via Doppler tomography (WASP-33b \citealt{Collier:2010b}, Kepler-13b \citealt{Szabo:2011,Shporer:2011,Johnson:2014}, HAT-P-57b \citealt{Hartman:2015}, and KELT-17b \citealt{2016arXiv160703512Z}). 

In this paper, we present the discovery of \hatcurb{}, a Saturn-mass planet found to transit an F-subgiant by the HATNet survey \citep{Bakos:2004}. Despite the evolved status of \hatcur{}, the host star still exhibits a rapid rotation rate of $\vsini = \genevastarvsini\,\kms$, making precise radial velocities difficult to obtain. Eventual confirmation was achieved via a detection of the Doppler tomographic shadow of the planet during transit. When a planet transits a rapidly rotating star, it successively blocks parts of the rotating stellar disk, causing an asymmetry in the observed spectral line profiles. At low rotational velocities, the asymmetry can be measured by the Holt-Rossiter-McLaughlin effect \citep{1893AstAp..12..646H,Rossiter:1924,McLaughlin:1924}. At higher rotational velocities, the shadow of the planet can be resolved in the broadened stellar spectroscopic lines \citep[e.g.][]{Collier:2010a,Collier:2010b}. A detection of the Doppler tomographic signal, at a depth and width that are in agreement with the photometric transit, eliminates eclipsing binary blend scenarios that may mimic transiting planet signals. Further radial velocity measurements can then provide an upper-limit mass constraint of the orbiting companion. If the mass can be constrained to less than that of brown dwarfs, the transiting object is confirmed to be a planet.


\section{Observations}
\label{sec:obs}

\subsection{Photometry}
\label{sec:detection}

\begin{figure}[!ht]
\plotone{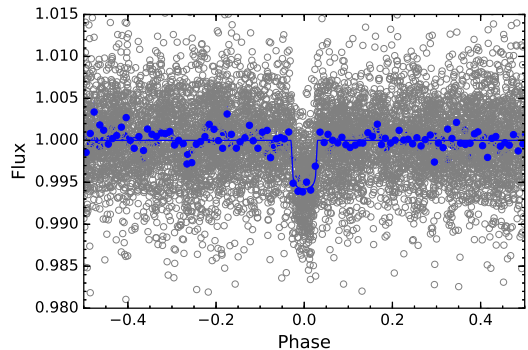}
\caption{
HATNet discovery light curves showing the transit of \hatcurb{}. The light curve is phase folded to a period of $P = \hatcurLCPprec$\,days, as per the analysis in Section~\ref{sec:analysis}. Grey points show the raw light curve, while blue points show the data binned at 0.01 in phase. Solid blue line shows the best fit transit model from Section~\ref{sec:global-fit}. 
\label{fig:hat_discovery}}
\end{figure}

The transits of \hatcurb{} were first detected with the HATNet survey \citep{Bakos:2004}. HATNet employs a network of small, wide field telescopes located at the Fred Lawrence Whipple Observatory (FLWO) in Arizona, and the Mauna Kea Observatory (MKO) in Hawaii, USA, to photometrically monitor selected $8\times 8^\circ$ fields of the sky. A total of 4050 $I$ band observations were taken by HAT-5 and HAT-8 over 2005 Jan -- July, and an additional 4518 observations were obtained in the Cousins~$R$ band using HAT-5, HAT-7, and HAT-8 telescopes between 2008 Feb -- Aug. The data reduction follows \citet{2010ApJ...710.1724B}. Light curves were produced via aperture photometry, and detrended with External Parameter Decorrelation \citep[EPD,][]{2007ApJ...670..826B} and Trend Filtering Algorithm \citep[TFA,][]{Kovacs:2005}. The Box-Fitting Least Squares \citep[BLS,][]{Kovacs:2002} analysis revealed the periodic transits of the planet candidate. The discovery light curve of \hatcurb{} is shown in Figure~\ref{fig:hat_discovery}, and the photometry presented in Table~\ref{tab:lc_table}.

\begin{deluxetable*}{rrrrrrr}

\tablewidth{0pc}
\tabletypesize{\scriptsize}
\tablecaption{
        Differential photometry of \hatcur{}
    \label{tab:lc_table}
}
\tablehead{
    \multicolumn{1}{c}{BJD}          &
    \multicolumn{1}{c}{Mag (Raw)}\tablenotemark{a}             &
    \multicolumn{1}{c}{Mag (EPD)}            &
    \multicolumn{1}{c}{Mag (TFA)}            &
    \multicolumn{1}{c}{$\sigma$ Mag}      &
    \multicolumn{1}{c}{Instrument}         &
    \multicolumn{1}{c}{Filter}            \\
    &
    &
    &
    &
}
\startdata

2454521.99042 & 9.39859 & 9.72414 & 9.72271 & 0.00303& HATNet & R\\
2454521.99445 & 9.39572 & 9.72048 & 9.73067 & 0.00325& HATNet & R\\
2454521.99854 & 9.39262 & 9.72005 & 9.71695 & 0.00276& HATNet & R\\
2454522.00264 & 9.38941 & 9.7199 & 9.72327 & 0.00344& HATNet & R\\
2454522.00673 & 9.41475 & 9.74011 & 9.73327 & 0.00328& HATNet & R\\
\enddata 
\tablenotetext{a}{
This table is available in a machine-readable form in the online journal. A portion is shown here for guidance regarding its form and content.\\
Raw, EPD, and TFA magnitudes are presented for HATNet light curves. The detrending and potential blending may cause the HATNet transit to be shallower than the true transit in the EPD and TFA light curves. This is accounted for in the global modelling by the inclusion of a third light factor. Follow-up light curves have been treated with EPD simultaneous to the transit fitting. Pre-EPD magnitudes are presented for the follow-up light curves.
}

\end{deluxetable*}

\begin{figure}[!ht]
\plotone{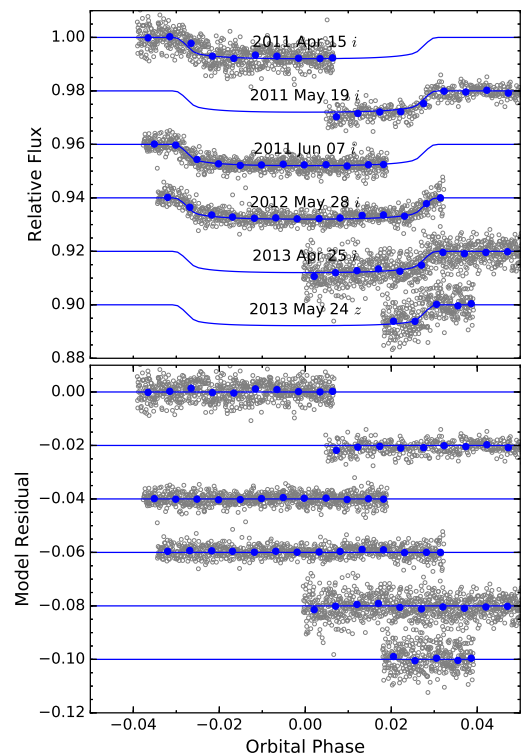}

\caption{
    Follow-up transit light curves of \hatcurb{} obtained by KeplerCam on the FLWO 1.2\,m telescope. The individual transits are labelled, and arbitrarily offset along the $y$ axis for clarity. The raw light curves are plotted in grey, and phase binned at 0.005 intervals in blue. The best fit models are plotted in blue. The residuals are shown on the bottom panel. 
\label{fig:fulc}} \end{figure}

To better characterize the planetary properties, follow-up photometry of the transits were obtained using KeplerCam on the FWLO 1.2\,m telescope. KeplerCam is a $4\mathrm{K} \times 4\mathrm{K}$ CCD camera with a pixel scale of $0\farcs672\,\mathrm{pixel}^{-1}$ at $2\times2$ pixel binning. The photometry was reduced as per \citet{2010ApJ...710.1724B}. A full transit was observed in the Sloan-$i$ band on 2012 May 28, and five partial transits observed on 2011 Apr 15, 2011 May 19, 2011 Jun 07, 2013 Apr 25 in the Sloan-$i$, and 2013 May 24 in the Sloan-$z$ band. The light curves and best fit models are shown in Figure~\ref{fig:fulc}, and the data presented in Table~\ref{tab:lc_table}.


\begin{deluxetable*}{llrrr}
\tablewidth{0pc}
\tabletypesize{\scriptsize}
\tablecaption{
    Summary of photometric observations
    \label{tab:photobs}
}
\tablehead{
    \multicolumn{1}{c}{Facility}          &
    \multicolumn{1}{c}{Date(s)}             &
    \multicolumn{1}{c}{Number of Images}\tablenotemark{a}      &
    \multicolumn{1}{c}{Cadence (s)}\tablenotemark{b}         &
    \multicolumn{1}{c}{Filter}            \\
    &
    &
    &
    &
}
\startdata
HATNet & 2005 Jan -- 2005 Jul & 4050 & 328 & $I$ \\
HATNet & 2008 Feb -- 2008 Aug & 4518 & 246 & Cousins~$R$ \\
FLWO 1.2\,m KeplerCam & 2011 Apr 15 & 730 & 24 & Sloan-$i$\\
FLWO 1.2\,m KeplerCam & 2011 May 19 & 509 & 44 & Sloan-$i$\\
FLWO 1.2\,m KeplerCam & 2011 Jun 07 & 801 & 29 & Sloan-$i$\\
FLWO 1.2\,m KeplerCam & 2012 May 28 & 730 & 34 & Sloan-$i$\\
FLWO 1.2\,m KeplerCam & 2013 Apr 25 & 960 & 24 & Sloan-$i$\\
FLWO 1.2\,m KeplerCam & 2013 May 24 & 361 & 24 & Sloan-$z$\\
\enddata 
\tablenotetext{a}{
  Outlying exposures have been discarded.
}
\tablenotetext{b}{
  Median time difference between points in the \lc. Uniform sampling was not possible due to visibility, weather, pauses.
}
\end{deluxetable*}

\subsection{Spectroscopy}
\label{sec:spec}

Spectroscopic observations of \hatcur{} were carried out using the FIber-fed Echelle Spectrograph (FIES), the Tillinghast Reflector Echelle Spectrograph (TRES), and the High Resolution Echelle Spectrometer (HIRES). The observations are summarized in Table~\ref{tab:specobssummary} and described below.

Initial spectroscopic characterization of \hatcurb{} was obtained with the FIES instrument \citep{2014AN....335...41T} on the 2.5\,m Nordic Optical Telescope (NOT). FIES is a fiber fed high resolution echelle spectrograph with a resolution of $ \lambda/\Delta \lambda \equiv R = 67000$ and spectral coverage of $3700-7300\,\AA$. Four FIES radial velocities were obtained over the 2009 Aug -- Oct period. The observations were obtained and reduced as per the procedure from \citet{Buchave:2010}. No radial velocity variation was detected with the FIES observations, with a scatter of $200\,\mathrm{m\,s}^{-1}$ over the four observations.

Additional observations were obtained with the TRES instrument \citep{Furesz:2008} on the FLWO 1.5\,m telescope. TRES is a fiber fed echelle with a spectral resolution of $R = 44000$, over the spectral region of $3850-9100\,\AA$. Radial velocities and spectral classifications are measured from each spectrum as per \citet{Buchave:2012}. Each TRES observation consists of three exposures combined together for cosmic-ray removal, and wavelength-calibrated by Th-Ar lamp exposures that bracket each set of three exposures. Two TRES observations at phase quadrature were taken in 2011 Apr 17 and 2011 Apr 20, with signal-to-noise at the Mg b lines of $\sim 100$ per resolution element. The velocity difference between the two observations was $80\,\mathrm{m\,s}^{-1}$, with a per-point uncertainty of $100\,\mathrm{m\,s}^{-1}$. As such, the the FIES and TRES observations showed that any companion orbiting \hatcur{} must be sub-brown dwarf in mass. 

In addition, we observed two partial spectroscopic transits of \hatcurb{}, on 2016 Apr 17 and 2016 May 16, with TRES to detect the Doppler tomographic shadow of the planet. These observations were performed as per the strategy described in \citet{Zhou:2016}. A set of time series spectra, at 900\,s cadence, where collected on both nights. The Doppler tomographic analysis for these two transit sets are described in Section~\ref{sec:TRES_DT}. 

\begin{figure} [ht]
\plotone{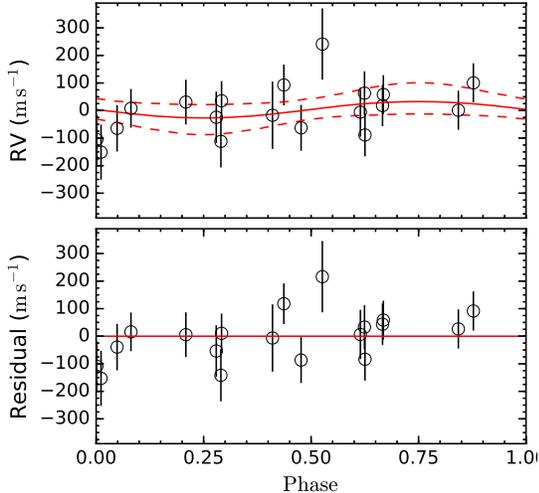}
\caption{
    Radial velocities from KECK-HIRES for \hatcur{}. The observations are marked by the open circles. The best fit circular orbit model is shown by the solid red line; the dashed lines encompass the $2\sigma$ set of models allowed by the data. The residuals are plotted on the bottom panel.
\label{fig:rv}}
\end{figure}

To constrain the mass of the companion, we obtained spectroscopic observations from HIRES on the 10\,m KECK telescope \citep{1994SPIE.2198..362V} at MKO over the 2009 Jul -- 2012 Mar period. A total of 19 observations were obtained through the $I_2$ cell to provide precise radial velocities. An additional $I_2$-free observation was obtained to provide a template for the radial-velocity measurements. The instrument was set up to use the C2 decker, which provides a $14\arcsec \times 0\farcs 861$ slit, yielding a spectral resolution of $R=48000$. The radial velocities were measured as per \citet{1996PASP..108..500B}, and the bisector spans calculated as per \citet{2007ApJ...666L.121T}. The high signal-to-noise HIRES observations provide the best constraints on the radial velocities of \hatcur{}, and were used in the global analysis in Section~\ref{sec:global-fit}. The radial velocities from HIRES are plotted in Figure~\ref{fig:rv} and presented in Table~\ref{tab:rv_table}. 

\begin{deluxetable*}{rrrrr}
\tablewidth{0pc}
\tabletypesize{\scriptsize}
\tablecaption{
       KECK-HIRES relative radial velocities and bisector span measurements of \hatcur{}
    \label{tab:rv_table}
}
\tablehead{
    \multicolumn{1}{c}{BJD}          &
    \multicolumn{1}{c}{RV}\tablenotemark{a}             &
    \multicolumn{1}{c}{$\sigma$ RV}      &
    \multicolumn{1}{c}{BS}         &
    \multicolumn{1}{c}{$\sigma$ BS}            \\
    \multicolumn{1}{c}{(UTC)} &
    \multicolumn{1}{c}{$(\mathrm{m\,s}^{-1})$} &
    \multicolumn{1}{c}{$(\mathrm{m\,s}^{-1})$} &
    \multicolumn{1}{c}{$(\mathrm{m\,s}^{-1})$} &
    \multicolumn{1}{c}{$(\mathrm{m\,s}^{-1})$}
}
\startdata
2455696.8366 & -105 & 28 & 24 & 11 \\ 
 2455696.88382 & -151 & 31 & 67 & 17 \\ 
 2455697.833 & 31 & 25 & -58 & 9 \\ 
 2455698.92918 & 93 & 23 & -53 & 8 \\ 
 2455699.83162 & 63 & 25 & 46 & 10 \\ 
 2455700.88206 & 1 & 22 & 30 & 13 \\ 
 2455704.84352 & 18 & 23 & -2 & 13 \\ 
 2455705.86007 & 100 & 22 & -34 & 8 \\ 
 2455706.83882 & 8 & 22 & 14 & 11 \\ 
 2455707.85238 & 35 & 22 & -17 & 13 \\ 
 2455853.70871 & -6 & 28 & ... & ... \\ 
2455945.15236 & -89 & 24 & ... & ... \\ 
2455997.02884 & -17 & 38 & ... & ... \\ 
2455017.0082 & 58 & 22 & ... & ... \\ 
2455042.88956 & -65 & 26 & ... & ... \\ 
2455043.9989 & -25 & 29 & ... & ... \\ 
2455044.94822 & -63 & 26 & ... & ... \\ 
2455048.86097 & -112 & 30 & ... & ... \\ 
2455107.71733 & 241 & 40 & -3 & 28 \\ 
\enddata 
\tablenotetext{a}{
 Internal errors excluding the component of astrophysical/instrumental jitter considered in Section 3. Bisector spans (BS) are given where available.
}

\end{deluxetable*}

%

\begin{deluxetable*}{llrrrrr}
\tablewidth{0pc}
\tabletypesize{\scriptsize}
\tablecaption{
    Summary of spectroscopic observations\label{tab:specobssummary}
}
\tablehead{
    \multicolumn{1}{c}{Telescope/Instrument} &
    \multicolumn{1}{c}{Date Range}          &
    \multicolumn{1}{c}{Number of Observations} &
    \multicolumn{1}{c}{Resolution}          &
    \multicolumn{1}{c}{Observing Mode}          \\
}
\startdata
NOT 2.5\,m/FIES & 2009 Aug 4 -- 2009 Oct 10 & 5 & 67000 & RECON RV\\
FLWO 1.5\,m/TRES & 2011 Apr 17 -- 2011 Apr 20 & 2 & 44000 & RECON RV\\
KECK 10\,m/HIRES & 2009 Jul 04 -- 2012 Mar 10 & 19 & 55000 & RV\tablenotemark{a} \\
FLWO 1.5\,m/TRES & 2016 Apr 17 & 14 & 44000 & Transit\tablenotemark{b} \\
FLWO 1.5\,m/TRES & 2016 May 16 & 16 & 44000 & Transit\tablenotemark{b} \\
\enddata 
\tablenotetext{a}{
  High resolution spectra to obtain stellar atmospheric parameters and high precision radial velocities
}
\tablenotetext{b}{
  High resolution in-transit spectra to detect the Doppler tomographic signal of the planet
}

\end{deluxetable*}

\section{Analysis}
\label{sec:analysis}

\subsection{Doppler tomographic detection of the planetary transit}
\label{sec:TRES_DT}

\begin{figure*}
\plottwo{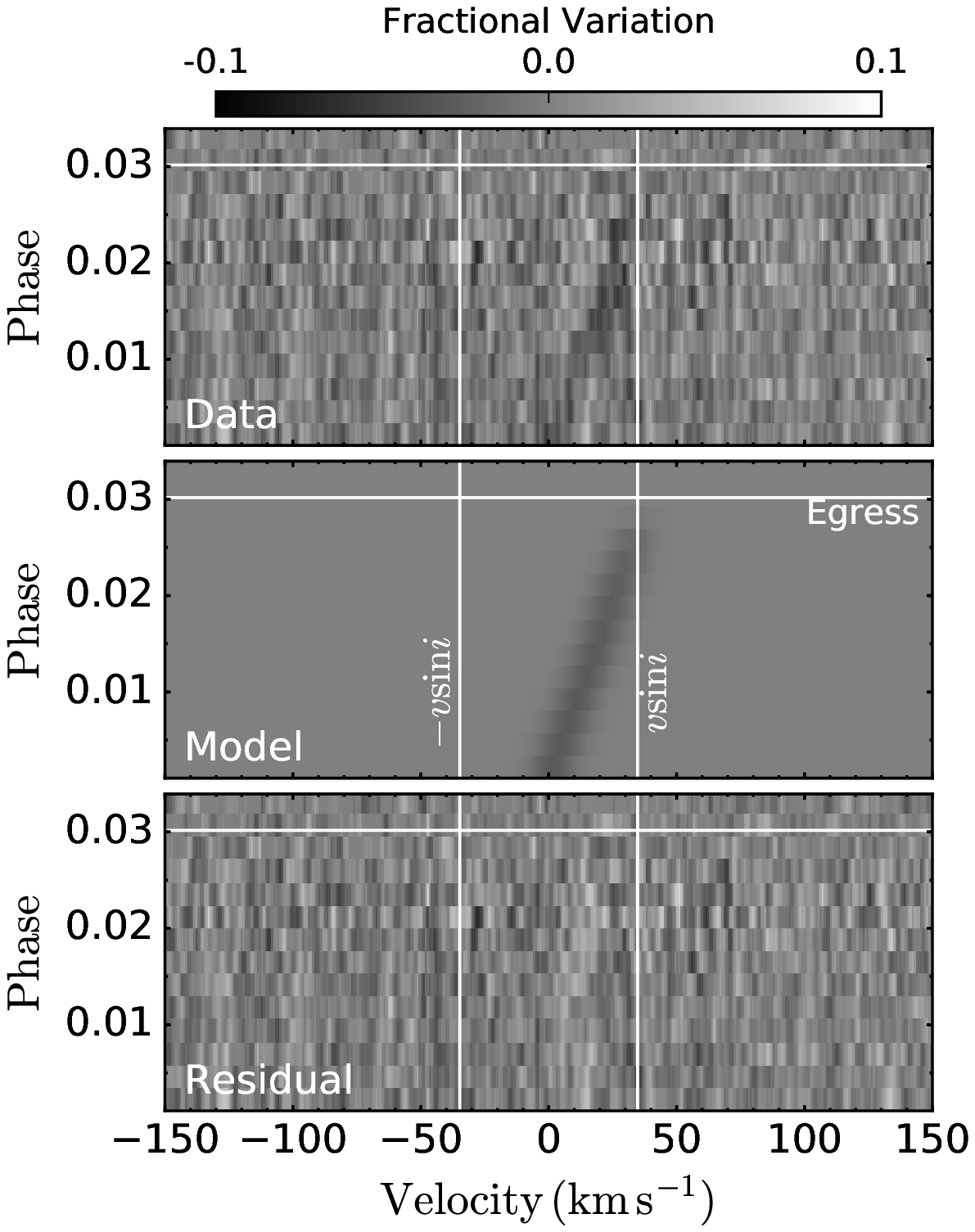}{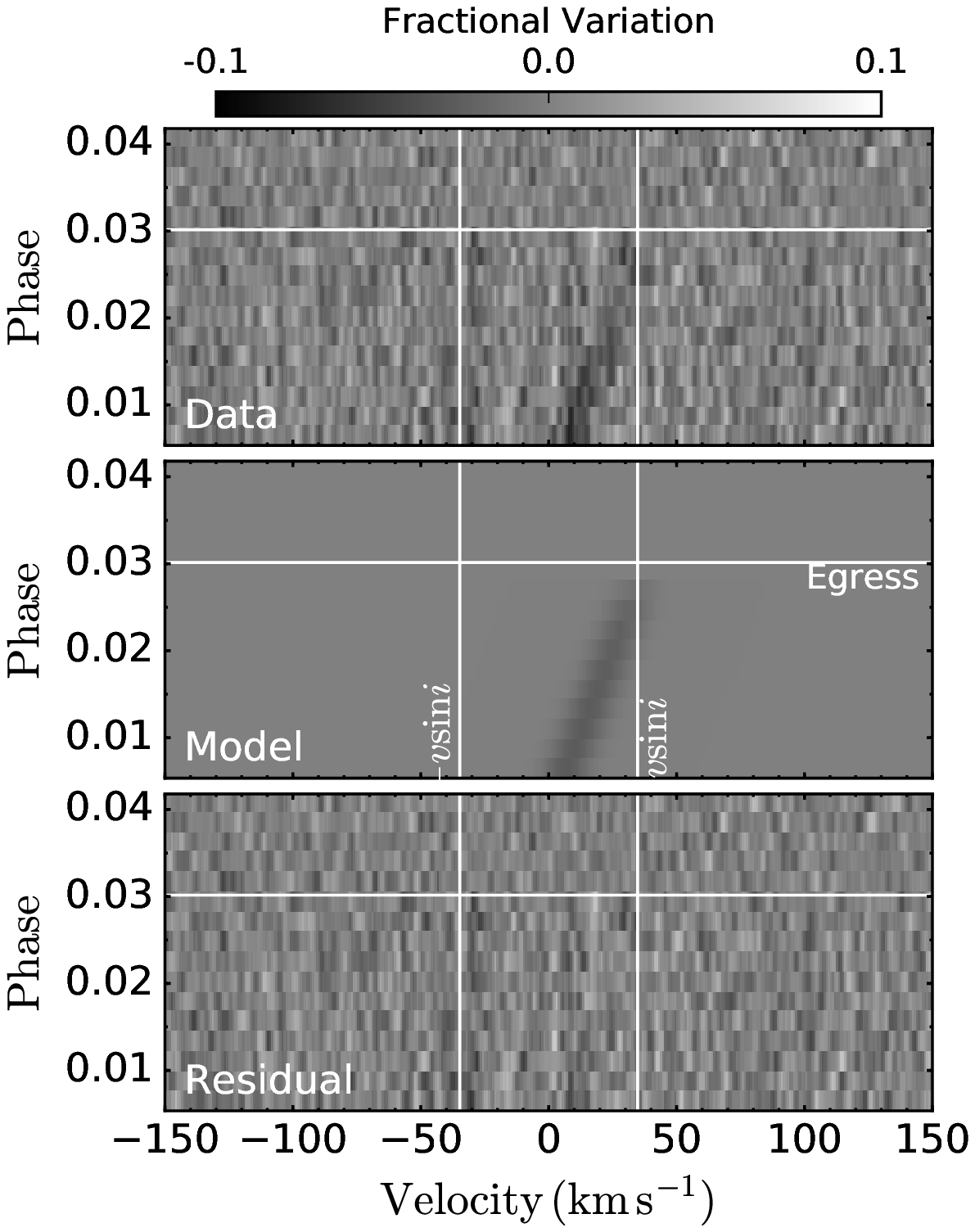}\\
\plottwo{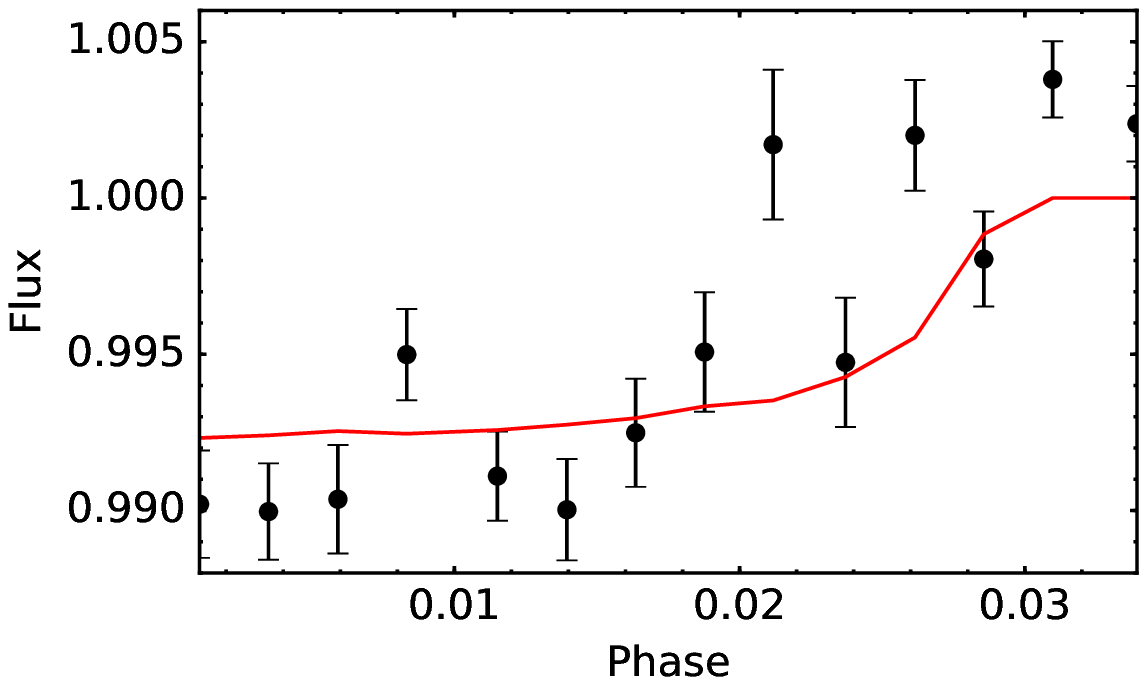}{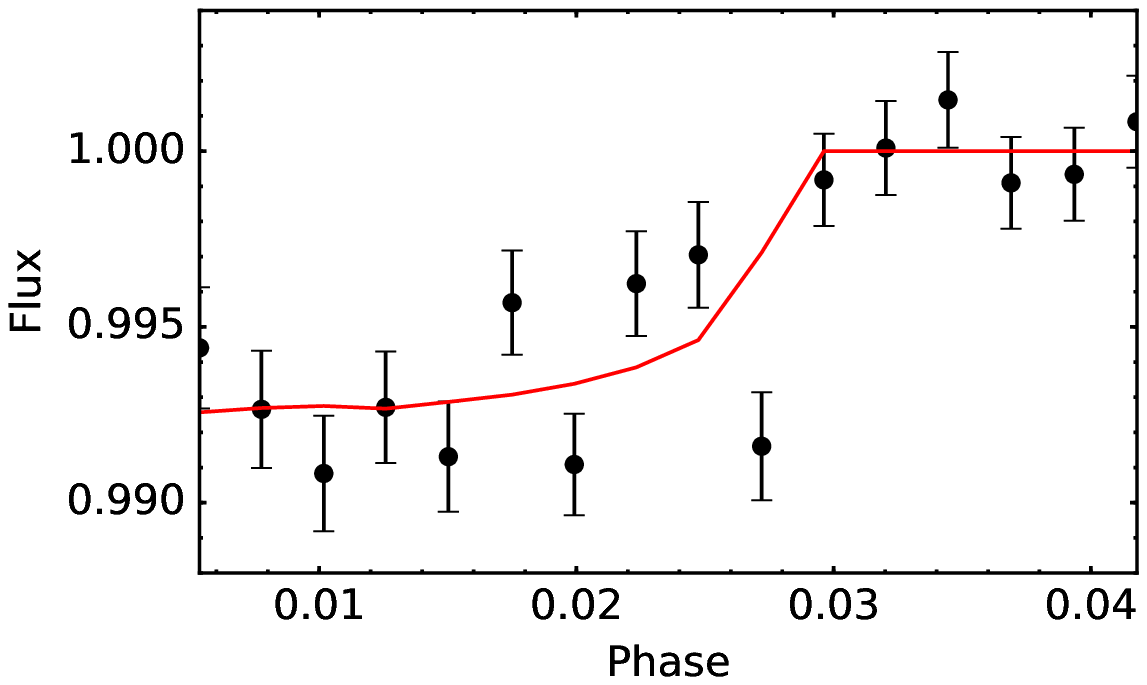}

\caption{
Doppler tomographic signals for the spectroscopic transits of \hatcurb{} on 2016 Apr 17 (left) and 2016 May 16 (right). The top panels show the residual between the broadening kernel from each observation and that of the averaged out-of-transit broadening kernel. The transit can be seen as the dark streak running diagonally from bottom left (mid-transit) to top right (post-egress). The best fit models are plotted below, as are the residual after subtraction of the modeled planetary tomographic signal. The bottom panels show the reconstructed light curves from the Doppler tomographic observation. These are constructed by summing the signal under the Doppler tomographic `shadow' of the planet. The red line shows the expected signal from the photometric transit, agreeing with the transit depth modelled via Doppler tomography, eliminating potential blend scenarios for the system.
\label{fig:dopplertomography}}
\end{figure*}

The significant rotational broadening of \hatcur{} allows us to detect the spectroscopic transit of the planet via Doppler tomography \citep{Collier:2010a,Collier:2010b}. Two sets of transit spectroscopy were obtained for \hatcurb{} with TRES. The TRES spectra were processed as per the procedure laid out in \citet{Zhou:2016}: the broadening profiles were derived via a least-squares deconvolution (LSD) of the observed spectra against a non-rotating stellar template \citep[as per ][]{Donati:1997}. Synthetic template spectra were generated using the SPECTRUM \citep{Gray:1994} spectral synthesis program, using the ATLAS9 model atmospheres \citep{Castelli:2004}. The synthetic templates were generated at the same \teff{}, \logg{} and \feh{} as \hatcur{}, with no line broadening imposed. A broadening profile was derived for each spectrum and subtracted from the average out-of-transit profile, revealing the planetary transit signal (Figure~\ref{fig:dopplertomography}). We model the rotational profiles and the planetary signal as part of our global analysis, described in Section~\ref{sec:global-fit}.

\subsection{Stellar parameters}
\label{sec:stel_params}
 Stellar atmospheric parameters of \hatcur{} were derived from the 32 TRES spectra using the Stellar Parameter Classification pipeline \citep[SPC,][]{Buchave:2012}. We first run SPC to retrieve an initial estimate of the stellar atmospheric parameters. These are then incorporated in a first run of the global modeling and isochrone retrieval analysis described later in Section~\ref{sec:global-fit}. We then re-run SPC with the stellar surface gravity $\logg$ fixed to that measured from the transit duration in the global analysis (Section~\ref{sec:global-fit}, with $\logg = \genevastarlogg$) to provide updated, and better constrained $\teff$ and $\feh$ values. We find that \hatcur{} is consistent with an F-subgiant, of effective temperature $T_\mathrm{eff} = 6406 \pm 62\,\mathrm{K}$, metallicity $\mathrm{[m/H]} = -0.08 \pm 0.05$, and projected rotational velocity $v \sin I_\star = 36.5 \pm 0.3\,\mathrm{km\,s}^{-1}$. Similarly, running SPC on the four FIES spectra yield $T_\mathrm{eff} = 6380 \pm 50\,\mathrm{K}$, $\log g = 3.91 \pm 0.10$, $\mathrm{[m/H]} = -0.05 \pm 0.08$, $v\sin I_\star = 38\,\mathrm{km\,s}^{-1}$, consistent with the interpretation that \hatcur{} is an F subgiant. Since an accurate \vsini{} measurement is vital to correctly modeling the Doppler tomographic signal, we also use the set of time-series TRES spectra to measure the \vsini{} of \hatcur{}. Following \citet{Zhou:2016}, the broadening kernel for each spectrum is modeled by a rotational kernel, with width of \vsini{}, and a Gaussian kernel to account for macroturbulence and instrumental broadening, finding $\vsini =  30.9 \pm 2.0 \,\kms$, and macroturbulence of $9.22\pm0.5\,\mathrm{km\,s}^{-1}$. The uncertainties are estimated from the standard deviation scatter between exposures.  The difference between the \vsini{} measured via SPC and that from the rotational profile can be partially attributed to the inclusion of macroturbulence. 

\subsection{GAIA parallax}
\label{sec:gaia}
\hatcur{} is included in the Tycho-GAIA-Astrometric-Catalogue in the first data release (DR1) of GAIA \citep{2016arXiv160904303L}, which measured a parallax of $2.60 \pm 0.23$\,mas. Several literature investigations have pointed out a systematic under-estimation in the DR1 parallaxes, as per separate studies via eclipsing binaries \citep{2016ApJ...831L...6S}, close-by Cepheids \citep{2016arXiv160905175C}, asteroseismic distances \citep{2016arXiv161108776S}, and comparison with existing parallaxes of solar neighborhood stars \citep{2016ApJ...832L..18J}. We adopt the correction offered in \citet{2016ApJ...831L...6S} of $-0.325\pm 0.062$\,mas to the DR1 parallax of \hatcur{}, arriving at an adopted parallax value of $2.92 \pm 0.23$\,mas, and corresponding astrometric distance measurement of $342 \pm 27$\,pc. This parallax measurement is used to co-constrain the stellar parameters during the global modeling in Section~\ref{sec:global-fit}.

\subsection{Global fitting and derived planet parameters}
\label{sec:global-fit}

We perform a global analysis of the HATNet discovery light curves, follow-up transit light curves, KECK-HIRES $I_2$ radial velocities, and the TRES Doppler tomographic signal, co-constrained by stellar isochrones and the GAIA distance measurement. The transits are modelled according to \citet{2002ApJ...580L.171M}, with the transit shape defined by the transit centroid time $T_0$, star-planet distance $a/R_\star$, planet-star radius ratio $R_p/R_\star$, and transit inclination $i$. Individual quadratic limb darkening parameters are assigned to each light curve \citep[interpolated from][]{Claret:2011}, and fixed throughout the fitting. Separate dilution factors are allowed for the HATNet $I$ and $R_C$ band light curves to account for any distortions to the light curve shape from the TFA detrending process. The follow-up light curves are simultaneously detrended against instrumental parameters describing the $X$, $Y$ pixel centroids of the target star, background flux, and target airmass. The radial velocities are described by an arbitrary offset $\gamma$ and orbital semi-amplitude $K$. The orbital eccentricity parameters $e\cos\omega$ and $e\sin\omega$ are also included when eccentricity is allowed to vary. The Doppler tomographic signal is modeled as per \citet{2016arXiv160703512Z}, via a 2D integration of the stellar surface covered by the planet. The free parameters describing the Doppler tomography effect include the projected spin-orbit angle $\lambda$, and the projected rotational broadening velocity $v\sin I_\star$. Note that we do not account for the broadening of the planetary shadow due to the motion of the planet during an exposure; the blurring of the planetary shadow during an exposure ($2\,\kms$) is smaller than the width of the shadow ($7.2\,\kms$), but is not an  insignificant effect. We also allow the effective temperature $T_\mathrm{eff}$, metallicity [M/H], and the apparent $K$-band magnitude to be iterated, though heavily constrained about their spectroscopic and photometric values. At each step, we derive a stellar density $\rho_\star$ from the transit duration as per \citet{Seager:2003,Sozzetti:2007}, and query the stellar isochrones to derive a distance modulus. Isochrone interpolation is performed at each step using the gradient boosting regression algorithm implemented in \emph{scikit-learn}. This distance modulus is compared to the actual distance as measured from the GAIA parallax, with the difference applied as a penalty on the likelihood function. 

The rapid rotation rate of \hatcur{} can introduce a bias in the isochrone-derived parameters for the system. For stars with radiative envelopes, the convective core overshoot and mixing length parameters are different to that of non-rotating stellar models, with the overall effect of lengthening the main-sequence lifetime \citep[e.g.][]{2000A&A...361..101M}. We adopt the \hatcurisoshort{} 2D stellar evolution models \citep{2012A&A...537A.146E}, which account for the effects of rotation, for our analysis. For the isochrone fitting, we introduce the added dimension of equatorial velocity $v_{eq}$ into our interpolation. The $v_{eq}$ distribution is calculated from the measured $v\sin I_\star$ value, scaled by a uniform distribution of orientations sampled in $\cos I_\star$. 

To compare our Geneva isochrone results to fittings with more traditional 1D isochrones, we also present the results from analyses using the Dartmouth isochrones \citep{2008ApJS..178...89D}.

The parameter space is explored with a Markov chain Monte Carlo (MCMC) analysis, using the affine-invariant ensemble sampler \emph{emcee} \citep{ForemanMackey:2012}. The observations are fitted for twice, with the per-point uncertainties for each dataset inflated such that the reduced $\chi^2$ is at unity for the second run. A $\cos i$ prior is imposed on the transit inclination, while a Gaussian prior is imposed on $T_\mathrm{eff} = 6406\pm 64\,\mathrm{K}$, $\mathrm{[M/H]}=0.08\pm0.05$, and $\vsini = 30.9\pm 2.0\mathrm{km\,s}^{-1}$ based on the spectroscopic values outlined in Section~\ref{sec:stel_params}. We note that the derived posterior $\vsini$ (\genevastarvsini\,\kms) is offset with the prior by $\sim 2\sigma$. Resetting the prior to \genevastarvsini\,\kms did not change the system parameters significantly, with a derived $\lambda$ $1\sigma$ upper limit of $<11^\circ$ (compared to $<14^\circ$ from our adopted results). The $K$-band magnitude is also constrained by a Gaussian prior about its 2MASS value \citep{Skrutskie:2006}. The GAIA parallax is also heavily constrained by a Gaussian prior about our adopted value of $2.92 \pm 0.23$\,mas as described in Section~\ref{sec:stel_params}. A $\beta$ distribution prior is imposed on the eccentricity, following the prescription for short period planets set out in \citet{2013MNRAS.434L..51K}. Uniform priors are imposed on all other parameters. 
 
\begin{figure}[!ht]
\plotone{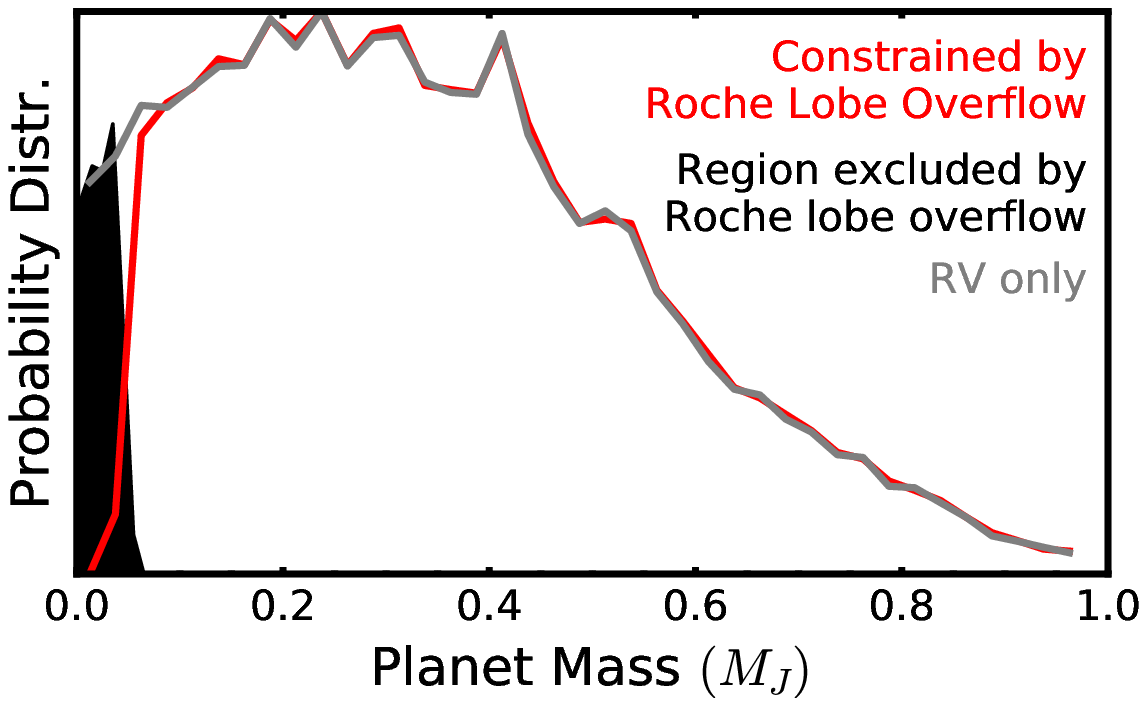}
\caption{
The posterior distribution for the mass of the planet. The grey line shows the posterior distribution constrained only by the radial velocities, from which an upper limit of $0.59\,M_J$ can be derived. A lower limit of $0.056\,M_J$ can also be applied if we assume the planet is not undergoing Roche lobe overflow. The resulting mass distribution is marked by the red line, while the solutions excluded are filled in black. 
\label{fig:massposterior}}
\end{figure}

Due to the large radius of \hatcurb{}, potential solutions in the MCMC chain lead to the planet overflowing its Roche lobe \citep[e.g.][]{2004A&A...418L...1L}. We can use this to place a lower limit on the mass of the planet by assuming no Roche lobe overflow. For each link of the MCMC chain, we calculate the corresponding Roche lobe radius using equation A5 of \citet{2011ApJ...742...59H}. Links with $R_p/a$ overflowing the Roche lobe are eliminated. For the circular orbit fit, the Roche lobe provides a weak lower limit on the mass of the planet of $0.056 \, M_\mathrm{J}$. The posterior distribution for planet mass is plotted in Figure~\ref{fig:massposterior}. The final mass measurement we report is the 68\% confidence interval for the Roche lobe constrained posterior distribution. 

We present four sets of solutions in Tables~\ref{tab:stellar} and \ref{tab:planetparam} for the circular and eccentric orbit scenarios from the Geneva and Dartmouth isochrone fits. The circular orbit Geneva isochrone fit solution is preferred, favored over the eccentric solution with a Bayesian Information Criterion $\Delta \mathrm{BIC}$ of 212. That is, the increased degrees of freedom in an eccentric orbit fit do not justify the improvements in the goodness of fit over that of a circular orbit model.

\begin{figure*}[!ht]
\plottwo{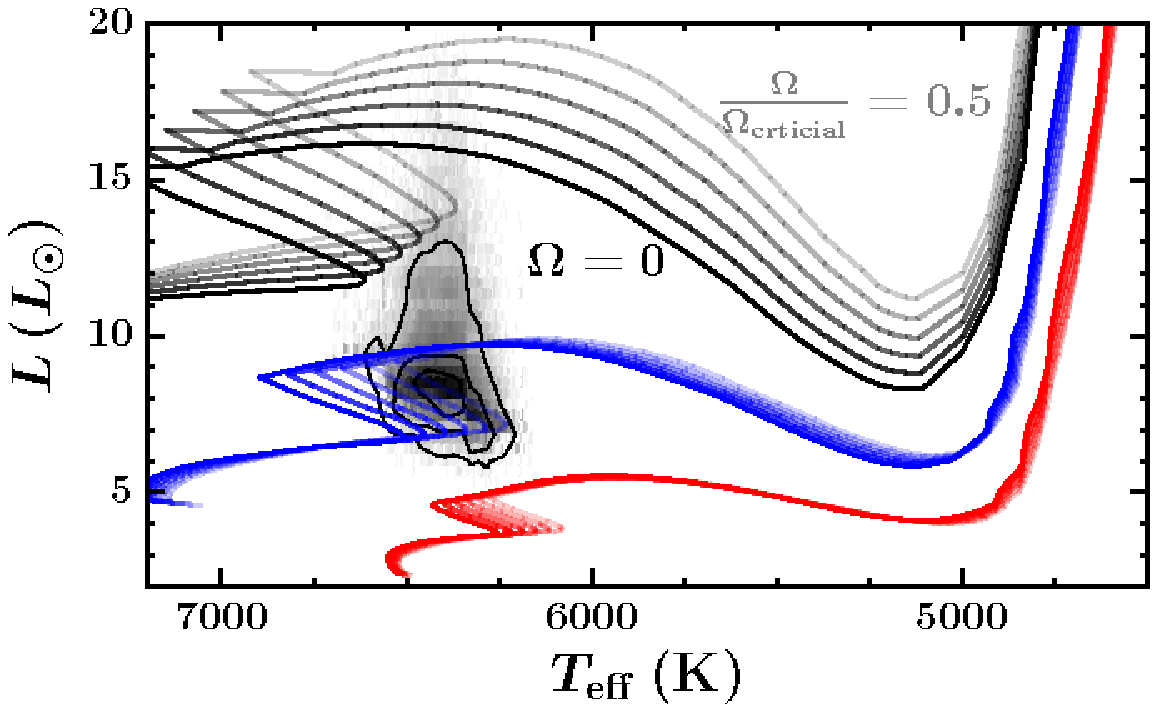}{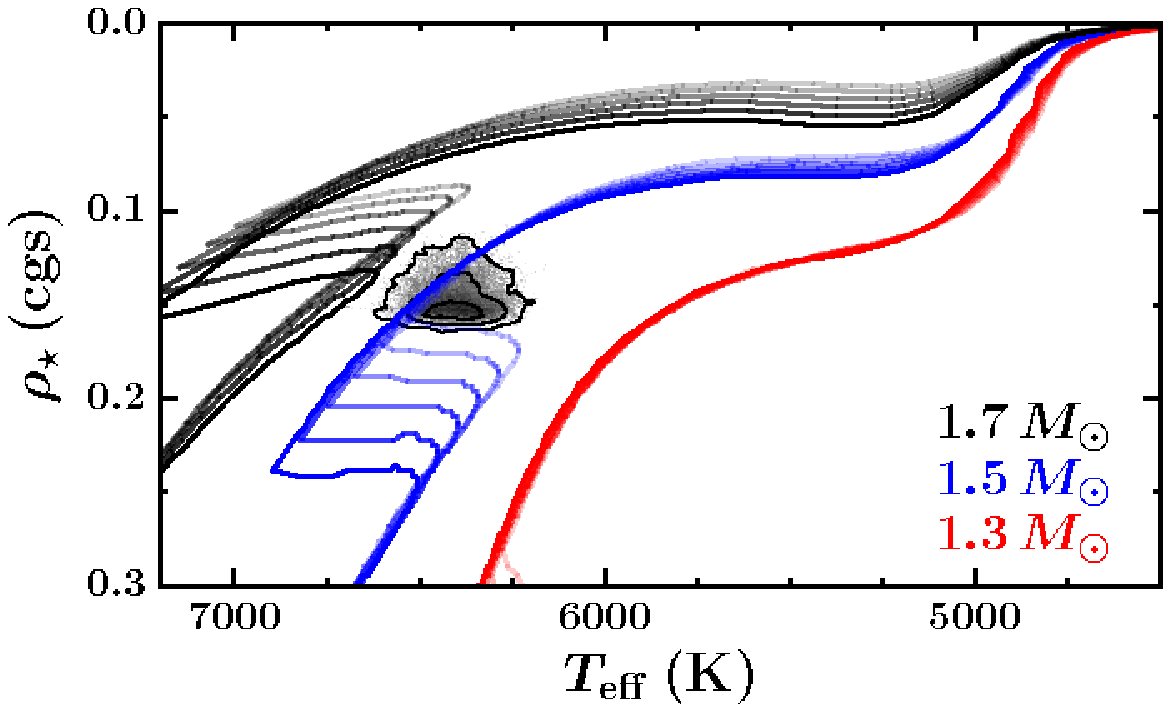}
\caption{
Model evolutionary tracks of effective temperature -- luminosity (\textbf{left}) and effective temperature -- stellar density (\textbf{right}) from the Geneva isochrones \citep{\hatcurisocite} are plotted for solar metallicity stars of various masses and rotation rates. Red tracks denote stars of 1.3\,\msun, blue for 1.5\,\msun, black for 1.7\,\msun. The shades of the lines illustrate the influence of rotation on evolution, with darkest for no rotation, and lightest for $\Omega / \Omega_\mathrm{critical} = 0.5$, at 0.1 intervals. The 1, 2, and $3\sigma$ contours for the posterior probability distribution of \hatcur{} are plotted. Note that the effective temperature -- stellar density distribution (\textbf{right}) is model independent, with effective temperature measured from spectra, and stellar density derived from the transit duration. The effective temperature -- luminosity distribution (\textbf{left}) requires isochrone interpolation of luminosity, and is therefore model dependent.  
\label{fig:iso}}
\end{figure*}

The evolutionary stage of \hatcur{} is shown in Figure~\ref{fig:iso} on the Hertzsprung-Russell diagram, along with evolutionary tracks of various stellar masses and rotation rates marked for context. The derived stellar and planetary parameters are presented in Tables~\ref{tab:stellar} and \ref{tab:planetparam}, respectively.


\begin{deluxetable*}{lrrrr}
\tablewidth{0pc}
\tabletypesize{\scriptsize}
\tablecaption{
    Stellar parameters for \hatcur{}
    \label{tab:stellar}
}
\tablehead{
    \multicolumn{1}{c}{~~~~~~~~Parameter~~~~~~~~}   &
    \multicolumn{1}{c}{\textbf{Circular Fit Geneva}} &
    \multicolumn{1}{c}{Eccentric Fit Geneva} &
    \multicolumn{1}{c}{Circular Fit Dartmouth} &
    \multicolumn{1}{c}{Eccentric Fit Dartmouth} 
}
\startdata
\noalign{\vskip -3pt}
\sidehead{Catalogue Information}
~~~~Tycho-2 \dotfill & 3084-533-1 &&& \\
~~~~GSC \dotfill & \hatcurCCgsc{} &&& \\
~~~~2MASS \dotfill& \hatcurCCtwomass{} &&& \\
~~~~GAIA \dotfill & 1358614978835493120 &&& \\
~~~~GAIA RA (J2015) \dotfill & 17:06:26.574 &&&\\
~~~~GAIA DEC (J2015) \dotfill& +44:46:36.794 &&& \\
~~~~GAIA $\mu_\alpha$ $(\mathrm{mas}\,\mathrm{yr}^{-1})$ \dotfill& $9.32 \pm 0.88$ &&& \\
~~~~GAIA $\mu_\delta$ $(\mathrm{mas}\,\mathrm{yr}^{-1})$ \dotfill& $18.5 \pm 1.2$  &&& \\
~~~~GAIA Parallax\tablenotemark{a} $(\mathrm{mas})$ \dotfill & $2.92 \pm 0.23$  &&& \\
\sidehead{Spectroscopic properties  \tablenotemark{b} \tablenotemark{c} }
~~~~$\teffstar$ (K)\dotfill       &  \genevastarteff & $6408_{-65}^{+63}$ & $6406_{-63}^{+58}$ & $6414_{-59}^{+69}$ \\
~~~~$\feh$\dotfill                &  $-0.08 \pm 0.05$ & $-0.08 \pm 0.05$ & $-0.07_{-0.05}^{+0.04}$ & $-0.08\pm0.05$ \\
~~~~$\vsini$ (\kms)\dotfill        &  \genevastarvsini & $35.8 \pm 1.1$ & $33.2_{-1.2}^{+1.5}$ & $33.9_{-1.3}^{+1.2}$\\
\sidehead{Photometric properties}
~~~~GALEX FUV (AB mag)\dotfill               & $19.759\pm 0.137$ &&& \\
~~~~GALEX NUV (AB mag)\dotfill               & $14.251\pm 0.007$ &&& \\
~~~~GAIA $g$ (mag)\dotfill               & 9.94 &&& \\
~~~~APASS $B$ (mag)\dotfill               &  \hatcurCCapassmBcirc &&&\\
~~~~APASS $g'$ (mag)\dotfill               &  $10.351$ &&&\\
~~~~APASS $V$ (mag)\dotfill               &  \hatcurCCapassmVcirc &&&\\
~~~~APASS $r'$ (mag)\dotfill               &  $10.010$ &&&\\
~~~~TASS $I$ (mag)\dotfill               &  $9.518\pm0.048$ &&&\\
~~~~2MASS $J$ (mag)\dotfill               &  \hatcurCCtwomassJmagcirc &&&\\
~~~~2MASS $H$ (mag)\dotfill               &  \hatcurCCtwomassHmagcirc &&&\\
~~~~2MASS $K_s$ (mag)\dotfill             &  \hatcurCCtwomassKmagcirc &&&\\
\sidehead{Derived properties\tablenotemark{b}}
~~~~$\mstar$ ($\msun$)\dotfill      & \genevastarmass &  $1.73_{-0.13}^{+0.21}$ & $1.43\pm0.05$ & $1.38_{-0.05}^{+0.05}$ \\
~~~~$\rstar$ ($\rsun$)\dotfill      & \genevastarradius &  $2.71_{-0.39}^{+0.48}$ & $2.389_{-0.038}^{+0.040}$ & $2.13_{-0.14}^{+0.17}$   \\
~~~~$\loggstar$ (cgs)\dotfill       & \genevastarlogg & $3.800_{-0.080}^{+0.106}$ & $3.837_{-0.011}^{+0.009}$ & $3.932_{-0.060}^{+0.035}$   \\
~~~~$\lstar$ ($\lsun$)\dotfill      & $8.68_{-0.86}^{+1.50}$ & $8.3_{-1.9}^{+4.0}$ & $8.62_{-0.50}^{+0.57}$ &  $6.8_{-0.9}^{+1.2}$  \\
~~~~$M_V$ (mag)\dotfill             & $2.50_{-0.23}^{+0.13}$ & $2.57_{-0.37}^{+0.29}$ & $2.403_{-0.063}^{+0.083}$ & $2.67_{-0.17}^{+0.15}$ \\
~~~~$M_K$ (mag,\hatcurjhkfilset)\dotfill & $1.26_{-0.34}^{+0.15}$ & $1.36_{-0.39}^{+0.25}$ & $1.304_{-0.045}^{+0.046}$ &  $1.56_{-0.18}^{+0.14}$   \\
~~~~$A_V$ (mag)\dotfill & $<0.051\,(1\sigma)$ &  $<0.061\,(1\sigma)$  & $<0.13\,(1\sigma)$ & $<0.11\,(1\sigma)$\\

~~~~Age (Gyr)\dotfill               & $1.24_{-0.22}^{+0.27}$ &  $1.00_{-0.41}^{+0.21}$ & $2.83_{-0.19}^{+0.22}$ &  $3.04_{-0.27}^{+0.31}$ \\
~~~~Distance (pc) \dotfill           & $320_{-14}^{+48}$ &   $322_{-19}^{+35}$ & $335_{-7}^{+7}$ & $297_{-18}^{+26}$ \\
\enddata
\tablenotetext{a}{
 A correction of $-0.325\pm 0.062$\,mas has been applied to the GAIA DR1 parallax, as per \citet{2016ApJ...831L...6S}.\\
}
\tablenotetext{b}{
  Derived from the global modelling described in Section~\ref{sec:global-fit}, co-constrained by spectroscopic stellar parameters and the GAIA parallax.\\
}
\tablenotetext{c}{
  These stellar parameters are heavily constrained by Gaussain priors about their derived values from the Keck-HIRES iodine-free spectrum using the Stellar Parameter Classification (SPC) pipeline \citep{Buchave:2012}.
}\end{deluxetable*}


\begin{deluxetable*}{lrrrr}
\tablewidth{0pc}
\tabletypesize{\scriptsize}
\tablecaption{
    Orbital and planetary parameters 
    \label{tab:planetparam}
}
\tablehead{
    \multicolumn{1}{c}{~~~~~~~~Parameter~~~~~~~~}   &
    \multicolumn{1}{c}{\textbf{Circular Fit Geneva}} &
    \multicolumn{1}{c}{Eccentric Fit Geneva} & 
    \multicolumn{1}{c}{Circular Fit Dartmouth} & 
    \multicolumn{1}{c}{Eccentric Fit Dartmouth} 
}
\startdata
\noalign{\vskip -3pt}
\sidehead{\Lc{} parameters}
~~~$P$ (days)             \dotfill    & $4.8101025_{-0.00000033}^{+0.00000043}$ & $4.8101038_{-0.00000037}^{+0.00000054}$ & $4.8101017_{-0.00000030}^{+0.00000034}$ & $4.8101082_{-0.00000051}^{+0.00000052}$ \\
~~~$T_c$ (${\rm BJD}$)    
      \tablenotemark{a}   \dotfill    & $2455961.38467_{-0.00064}^{+0.00076}$ & $2455961.38472_{-0.00082}^{+0.00090}$ & $2455961.38465_{-0.00065}^{+0.00074}$ & $2455961.3852_{-0.0010}^{+0.0008}$ \\
~~~$T_{14}$ (days)
      \tablenotemark{a}   \dotfill    & $0.2912 \pm 0.0019$ & $0.308_{-0.031}^{+0.029}$ & $0.2910\pm{0.0015}$ & $0.257_{-0.010}^{+0.019}$\\
~~~$T_{12} = T_{34}$ (days)
      \tablenotemark{a}   \dotfill    & $0.0229\pm0.0010$ & $0.0246\pm 0.0027$ & $0.02330_{-0.00055}^{+0.00030}$ & $0.0213_{-0.0014}^{+0.0015}$ \\
~~~$\arstar$              \dotfill    & $5.691_{-0.124}^{+0.057}$ &  $5.34_{-0.46}^{+0.61}$ & $5.659_{-0.061}^{+0.066}$ & $6.34_{-0.42}^{+0.30}$\\
~~~$\rpl/\rstar$          \dotfill    & $0.0834\pm0.0017$ &  $0.084_{-0.0020}^{+0.0019}$ & $0.0821_{-0.0009}^{+0.0013}$ & $0.0846_{-0.0018}^{+0.0016}$\\
~~~$b \equiv a \cos i/\rstar$
                          \dotfill    & $0.12_{-0.08}^{+0.12}$ & $0.12_{-0.08}^{+0.12}$ & $0.214_{-0.045}^{+0.023}$ & $0.20_{-0.12}^{+0.11}$ \\
~~~$i$ (deg)              \dotfill    & $88.8_{-1.3}^{+1.1}$ & $88.9\pm{1.6}$ & $88.37_{-0.57}^{+0.61}$ & $88.2_{-1.1}^{+1.3}$\\
~~~$|\lambda|$ (deg)      \dotfill    & \genevalambda $(<14\,1\sigma)$ & $2.5_{-4.6}^{+5.8}$ $(<12\,1\sigma)$ & $-1.6_{-4.6}^{+3.9}$ $(<4\,1\sigma)$ & $2.3_{-6.4}^{+6.6}$ $(<13\,1\sigma)$\\

\sidehead{Limb-darkening coefficients \tablenotemark{b}}
~~~$a_r$ (linear term)   \dotfill    & 0.2497 &&&\\
~~~$b_r$ (quadratic term) \dotfill    & 0.3765 &&&\\
~~~$a_I$                 \dotfill    & 0.1701  &&& \\
~~~$b_I$                 \dotfill    & 0.3744 &&&\\
~~~$a_i$                 \dotfill    & 0.1897 &&&\\
~~~$b_i$                  \dotfill    & 0.3747 &&& \\
~~~$a_z$                 \dotfill    & 0.1397 &&& \\
~~~$b_z$                  \dotfill    & 0.3661 &&&\\

\sidehead{RV parameters}
~~~$K$ (\ms)              \dotfill    & $<36\,(1\sigma)$  & $<52\,(1\sigma)$ & $<38\,(1\sigma)$ & $<37\,(1\sigma)$\\
~~~$e\cos\omega$ 
                          \dotfill    & & $-0.21_{-0.14}^{+0.15}$ & & $-0.03_{-0.22}^{+0.20}$\\
~~~$e\sin\omega$
                          \dotfill    & & $0.027_{-0.11}^{+0.10}$ && $-0.150_{-0.055}^{+0.075}$\\
~~~$e$                    \dotfill    & & $0.24\pm0.12$ && $0.22_{-0.08}^{+0.12}$ \\
~~~$\omega$               \dotfill    & & $172_{-43}^{+31}$ && $105_{-66}^{+46}$\\
~~~RV jitter (\ms)\tablenotemark{c}        
                          \dotfill    & 59 & 58 & 59 & 59 \\
~~~Systemic RV (\kms)\tablenotemark{d}        
                          \dotfill    & $-1.4 \pm 0.5$ &&&  \\

\sidehead{Planetary parameters}
~~~$\mpl$ ($\mjup$)\tablenotemark{e}       \dotfill    & \genevaplanetmass & $0.49_{-0.22}^{+0.27}$ & $0.33_{-0.17}^{+0.22}$ & $0.29_{-0.19}^{+0.24}$\\
~~~$\rpl$ ($\rjup$)       \dotfill    & \genevaplanetradius &  $2.25_{-0.23}^{+0.20}$ & $1.975_{-0.038}^{+0.045}$ & $1.78_{-0.10}^{+0.14}$ \\
~~~$\rhopl$ (\gcmc)       \dotfill    &  \genevaplrho & $0.058_{-0.025}^{+0.039}$ & $0.058_{-0.030}^{+0.039}$ & $0.065_{-0.044}^{+0.062}$\\
~~~$\log g_p$ (cgs)       \dotfill    & $2.32_{-0.34}^{+0.24} $ & $2.41_{-0.25}^{+0.20}$ & $3.837_{-0.011}^{+0.009}$ & $2.36_{-0.47}^{+0.28}$ \\
~~~$a$ (AU)               \dotfill    & $0.06505_{-0.00079}^{+0.00273} $ & $0.0663_{-0.0014}^{+0.0016}$ & $0.062844_{-0.00049}^{+0.00053}$ & $0.061994_{-0.00072}^{+0.00068}$ \\
~~~$T_{\rm eq}$ (K)       \dotfill    & $1903\pm25$ & $1963_{-99}^{+85}$ & $1903_{-21}^{+19}$ & $1803_{-43}^{+62}$\\
~~~$\Theta$\tablenotemark{f}\dotfill  & $0.0138_{-0.0075}^{+0.0099}$ & $0.0178_{-0.0077}^{+0.0098}$ & $0.015_{-0.015}^{+0.025}$ & $0.015_{-0.010}^{+0.013}$\\
~~~$\langle F \rangle$  ($10^9$\ergscmsq) \tablenotemark{g} & $2.74_{-0.17}^{+0.19}$ & $2.57_{-0.46}^{+0.45}$ & $2.98_{-0.13}^{+0.14}$ & $2.41_{-0.25}^{+0.33}$ \\
\enddata
\tablenotetext{a}{
    \ensuremath{T_c}: Reference epoch of mid transit that minimizes the
    correlation with the orbital period. BJD is calculated from UTC.
    \ensuremath{T_{14}}: total transit duration, time between first to
    last contact;
    \ensuremath{T_{12}=T_{34}}: ingress/egress time, time between first
    and second, or third and fourth contact.
}
\tablenotetext{b}{
        Values for a quadratic law given separately for each of the filters with which photometric observations were obtained.  These values were adopted from the
        tabulations by \citet{Claret:2011} according to the
        spectroscopic (SPC) parameters listed in \reftabl{stellar}. The limb darkening coefficients are held fixed during the global modelling.
}
\tablenotetext{c}{
    This jitter was added linearly to the RV uncertainties for
    each instrument such that $\chi^{2}/{\rm dof} = 1$ for the
    observations from that instrument.
}
\tablenotetext{d}{
    The systemic RV for the system as measured relative to the telluric lines 
}

\tablenotetext{e}{
    The mass measurement is quoted as the median of the posterior, with the uncertainties defined as the 68 percentile region.
}

\tablenotetext{f}{
    The Safronov number is given by $\Theta = \frac{1}{2}(V_{\rm
    esc}/V_{\rm orb})^2 = (a/\rpl)(\mpl / \mstar )$
    \citep[see][]{2007ApJ...671..861H}.
}
\tablenotetext{g}{
    Incoming flux per unit surface area, averaged over the orbit.
}
\end{deluxetable*}

\subsection{Eccentricity constraint}
\label{sec:ecc_constraint}

\begin{figure}[!ht]
\plotone{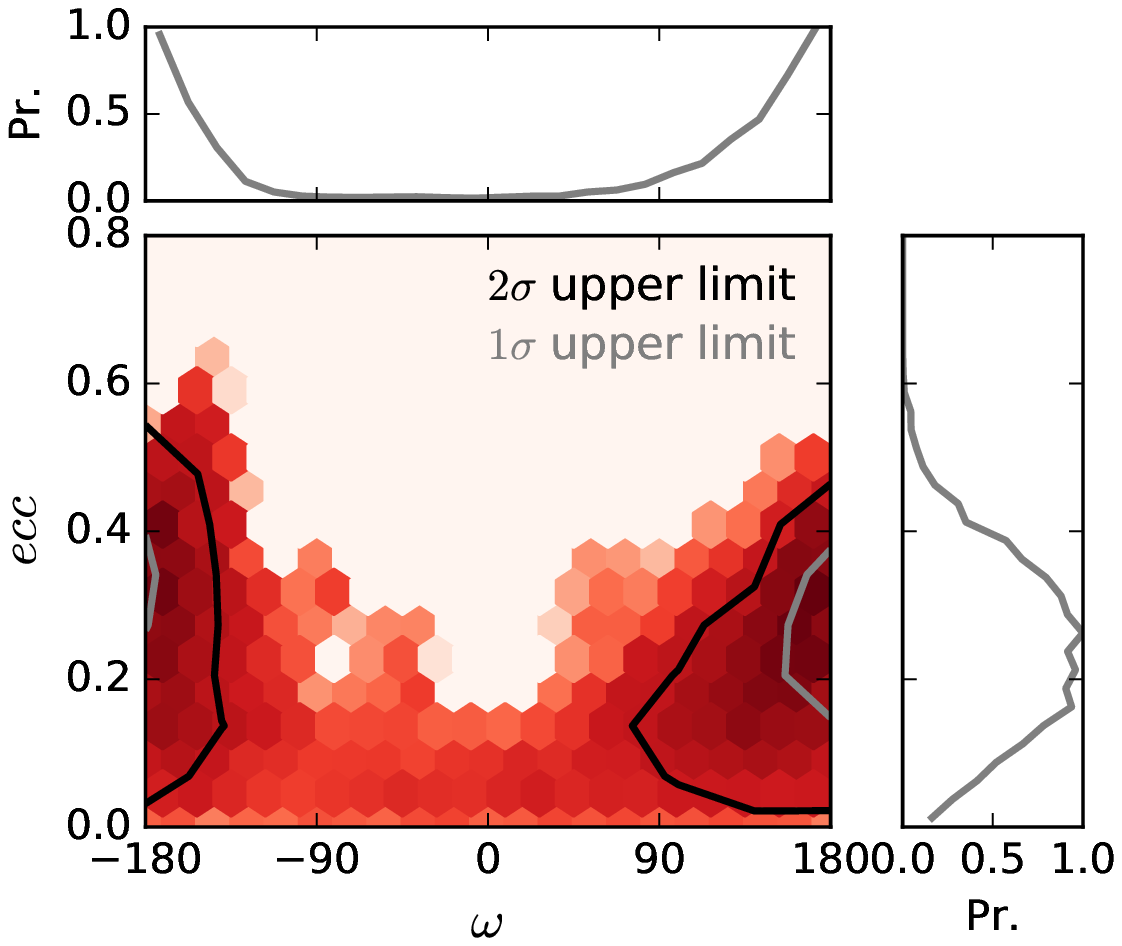}
\caption{
The eccentricity of \hatcurb{} is largely determined by the transit duration and the stellar radius derived from the light curves and GAIA distance. The eccentricity $ecc$ and argument of periastron $\omega$ posteriors are plotted. The 64 and 95 percentile contours are plotted in grey and black, respectively. 
\label{fig:eccposterior}}
\end{figure}

We can constrain the eccentricity of the system via the photometric light curves despite a lack of detection of the radial velocity orbit, since the GAIA parallax provides a good constraint on the stellar radius and transit duration \citep{2008MNRAS.389.1383K,2012ApJ...756..122D}. The eccentricity posterior, as constrained primarily from this `photo-eccentric' effect, is shown in Figure~\ref{fig:eccposterior}. The eccentricity $2\sigma$ upper limit is 0.43, with a posterior median and 64\% confidence region of $ecc = 0.24\pm0.12$.

The parallax we choose to adopt has an effect on our best-fit solutions. If we choose to adopt the GAIA parallax of $2.60 \pm 0.23$\,mas ($385 \pm 34$pc) from \citet{2016arXiv160904303L} without the systematic correction offered by \citet{2016ApJ...831L...6S}, we would have a modest $1.3\,\sigma$ tension between the best fit isochrone distance and the parallax distance. Adopting a distance of $385\pm34$pc yields an eccentric orbit of $e = 0.356_{-0.077}^{+0.072}$. The tidal circularization time scale for the system is $<500$\,Myr \citep{2004ApJ...610..464D}, so the likelihood of such an eccentric orbit is low for the system.

\subsection{Transit timing variations and additional companions}
\label{sec:ttv}

\begin{figure}[!ht]
\plotone{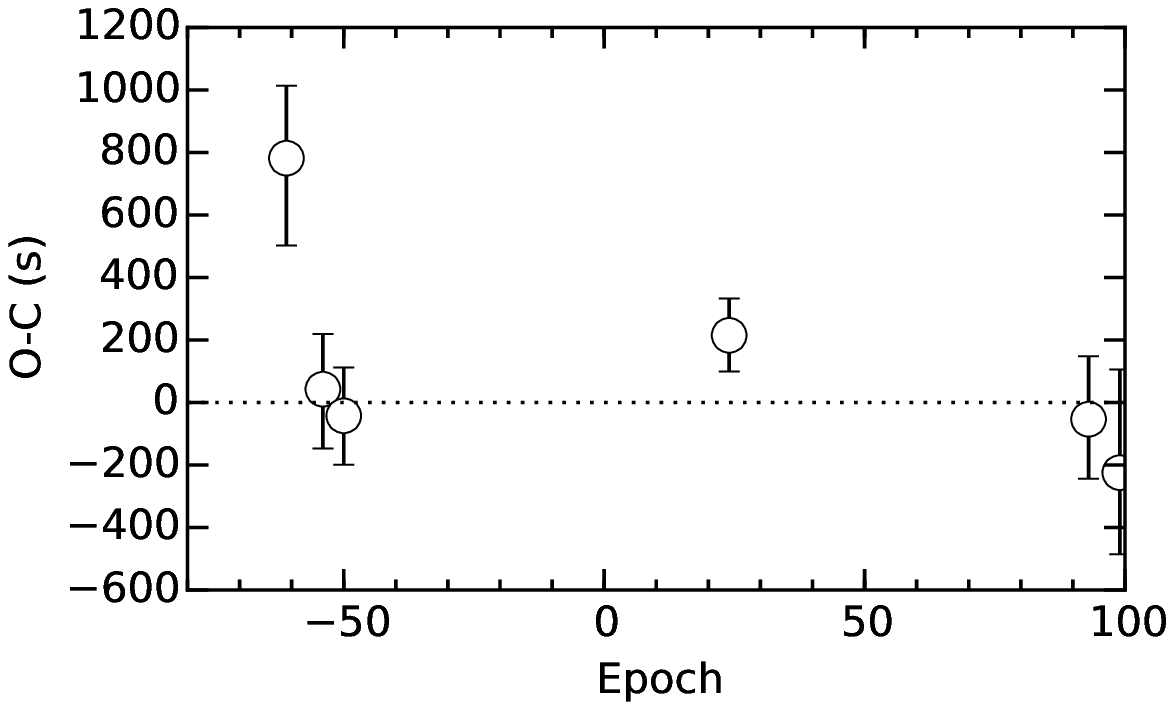}
\caption{
Transit centroid offsets $(O-C)$ for the follow-up light curves. We find no convincing evidence for transiting timing variations that may be indicative of additional orbiting companions. 
\label{fig:ttv}}
\end{figure}

To check for potential transit timing variations that may be indicative of additional orbiting companions, we re-fit the follow-up transit observations, allowing for individual transit centroids for each epoch. The timing residuals are shown in Figure~\ref{fig:ttv}. The transit geometry parameters $a/R_\star$, $R_p/R_\star$, and inclination, are heavily constrained by Gaussian priors about their best fit values from the global analysis (adopted as the circular orbit fit in Table~\ref{tab:planetparam}). We find no convincing evidence for transit timing variations, but also note that the $\sim 7$\,hr transit duration makes it difficult for us to capture full transits via ground-based follow-up, and partial transits provide poorer transit timing measurements. In addition, we find no evidence for long term radial velocity trend, with the quadratic and linear fits to the radial velocity data consistent with flat slopes. 

\subsection{Imaging Constraints on Resolved Neighbors}
\label{sec:image}

In order to detect possible neighboring stars which may be diluting
the photometric transits, we obtained optical and near infrared imaging of HAT-P-67 using the Clio2 near-IR imager \citep{freed:2004} on the MMT
6.5\,m telescope on Mt.~Hopkins, in AZ, together with the
Differential Speckle Survey Instrument
\citep[DSSI;][]{howell:2011,horch:2012,horch:2011} and the WIYN High-Resolution Infrared Camera (WHIRC), both on the WIYN~3.5\,m
telescope\footnote{The WIYN Observatory is a joint facility of the University of Wisconsin-Madison, Indiana University, the National Optical Astronomy Observatory and the University of Missouri.} at Kitt Peak National Observatory in Arizona.

%
%
\begin{figure*}[!ht]
\plottwo{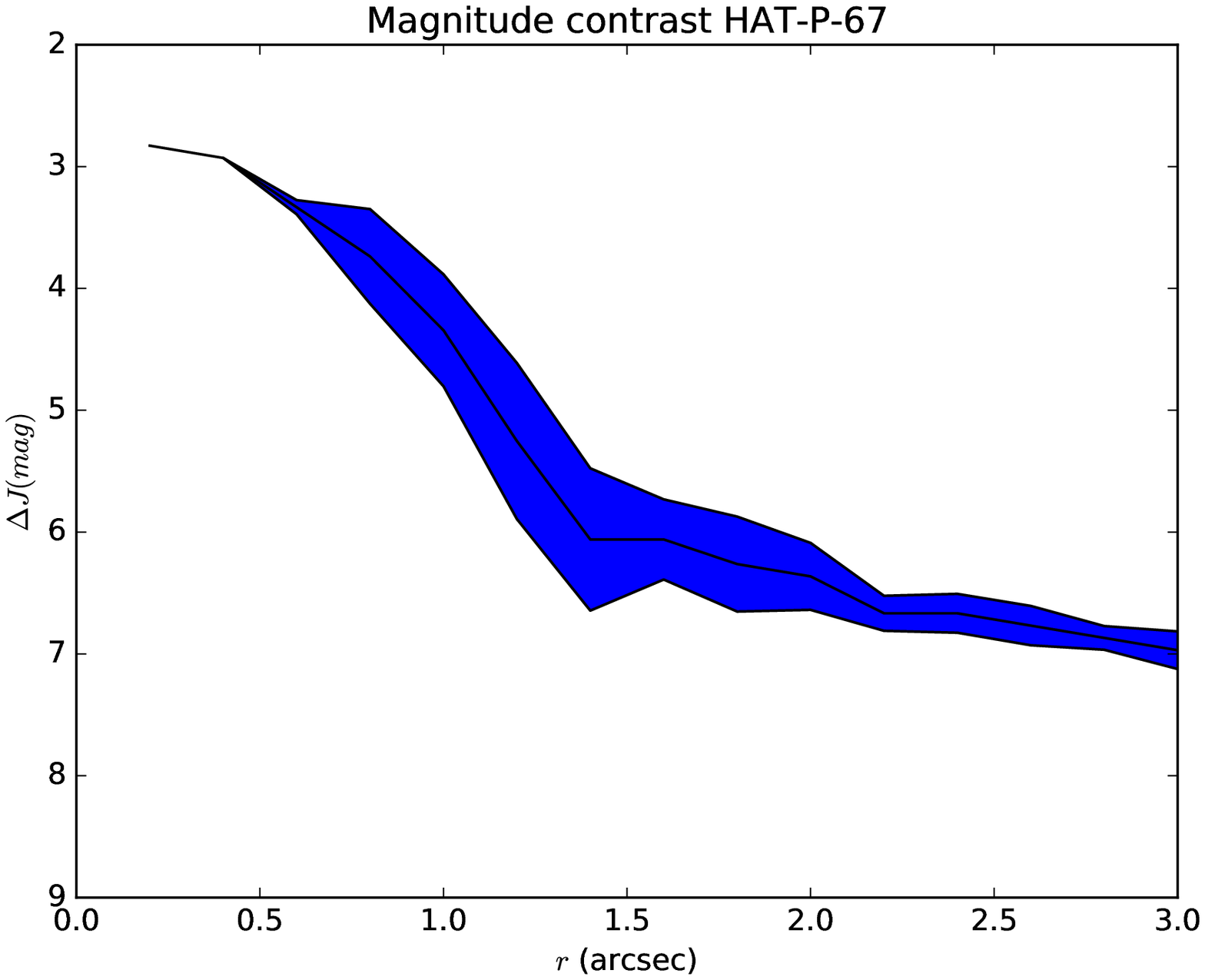}{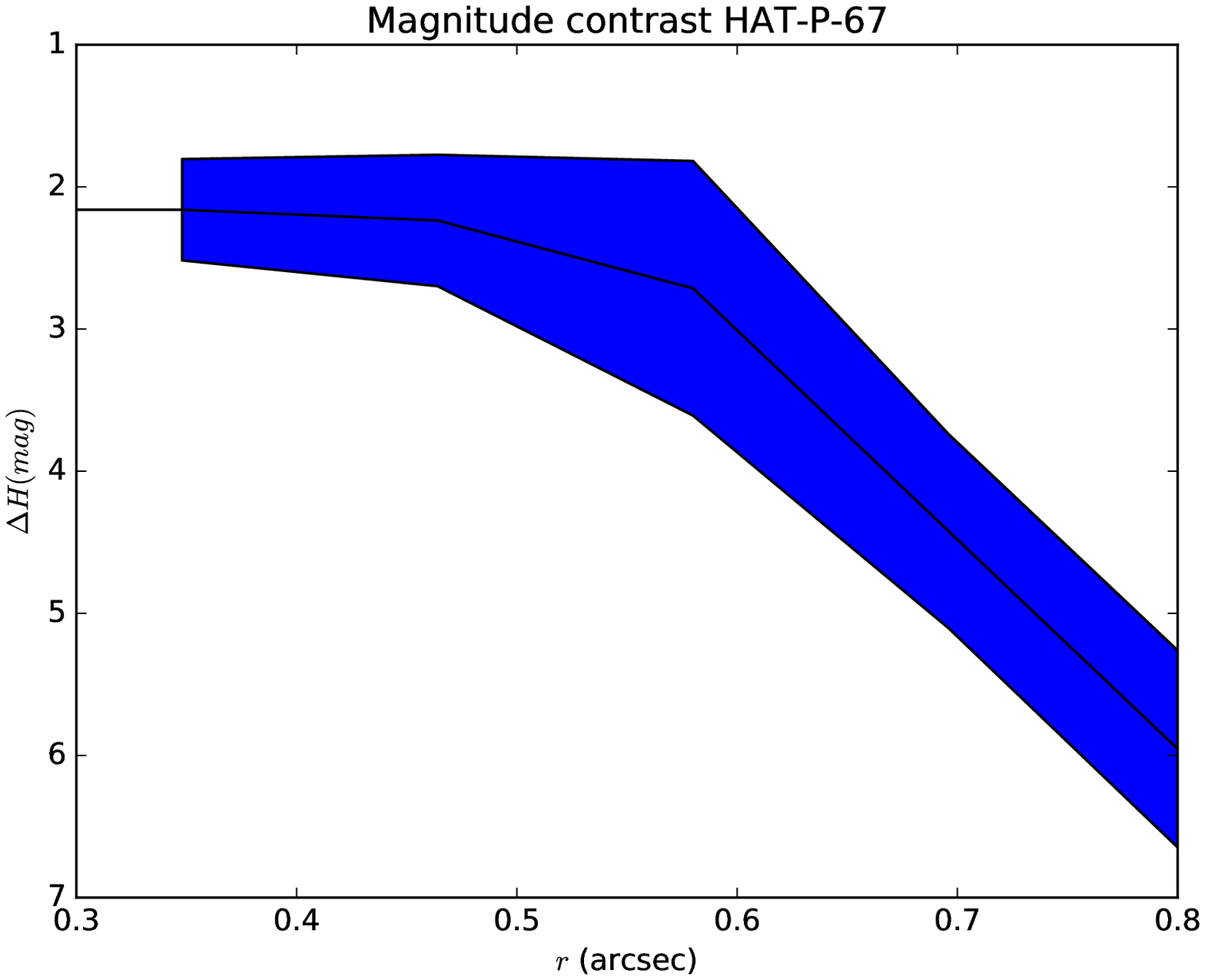}
\caption[]{
    Contrast curve for HAT-P-67 in
    the ({\em Left}) $J$-band based on observations
    made with WHIRC on the WIYN~3.5\,m and ({\em Right}) the $H$-band based on Clio2/MMT observations as described in
    Section~\ref{sec:image}. The bands show the variation in the
    contrast limits depending on the position angle of the putative
    neighbor.
\label{fig:whircclio2contrast}}
\end{figure*}

The Clio2 images were obtained on the night of UT
2011 June 22.  Observations in $H$-band and
$L^{\prime}$-band were made using the adaptive optics (AO)
system. A possible neighbor was detected $4\farcs9$ to the southeast of HAT-P-67 with a relative magnitude difference of $\Delta H = 7.4 \pm 0.5$\,mag, but no closer objects are seen.  The neighbor was blended with HAT-P-67 in the HATNet survey observations, but was fully resolved by all subsequent follow-up observations. Figure~\ref{fig:whircclio2contrast} shows the $H$-band magnitude contrast curve for HAT-P-67 based on these observations. This curve was calculated using the method and software
described by \citet{2016AJ....152..108E}. The band shown in this
image represents the variation in the contrast limit depending on the
position angle of the putative neighbor. We can rule out other neighbors with a magnitude difference of $\Delta H < 2$\,mag, down to a separation of $0\farcs3$, and $\Delta H < 6$\,mag, down to a separation of $0\farcs8$. The $L^{\prime}$ observations suffered from high thermal background, and the $4\farcs9$ neighbor was not detected. Meaningful constraints could not be placed on closer neighbors in $L^{\prime}$ based on these observations.

$J$-band snapshot images of HAT-P-67 were obtained with WHIRC on the night of 2016 April 24, with a seeing of $\sim 0\farcs9$. The images were collected at four nod positions, and were calibrated, background-subtracted, registered and median-combined using the same tools that we used for reducing the KeplerCam images. The $4\farcs9$ neighbor was not detected, and we concluded that it must have $\Delta J > 7$\,mag. The closest neighbor detected in these observations was at a separation of $9\farcs3$ to the northwest, and has a relative magnitude difference of $\Delta J = 4.96 \pm 0.01$\,mag compared to HAT-P-67. Figure~\ref{fig:whircclio2contrast} shows the $J$-band magnitude
contrast curve computed in a similar manner to the $H$-band contrast curve.

\begin{figure*}[!ht]
{
\plottwo{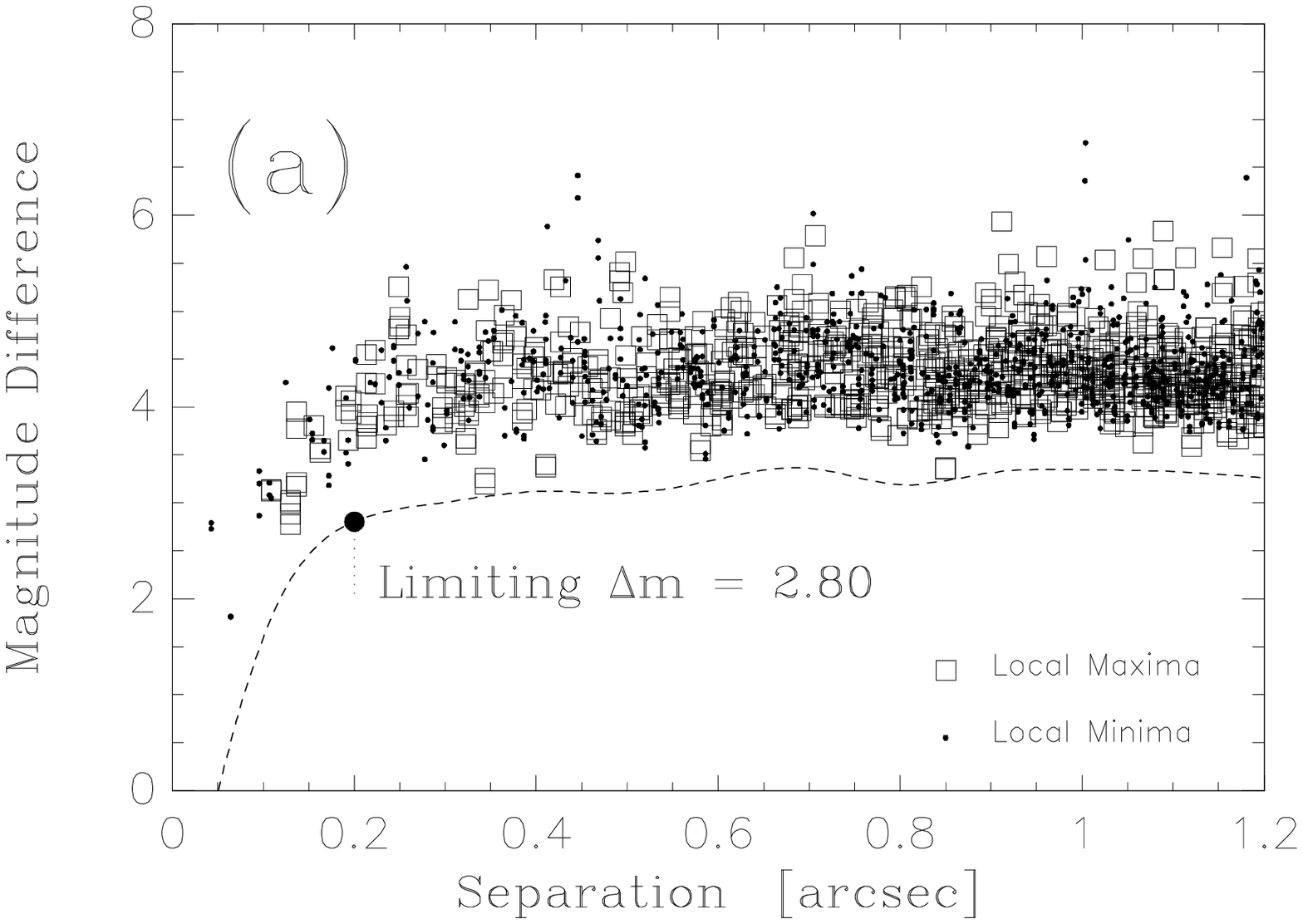}{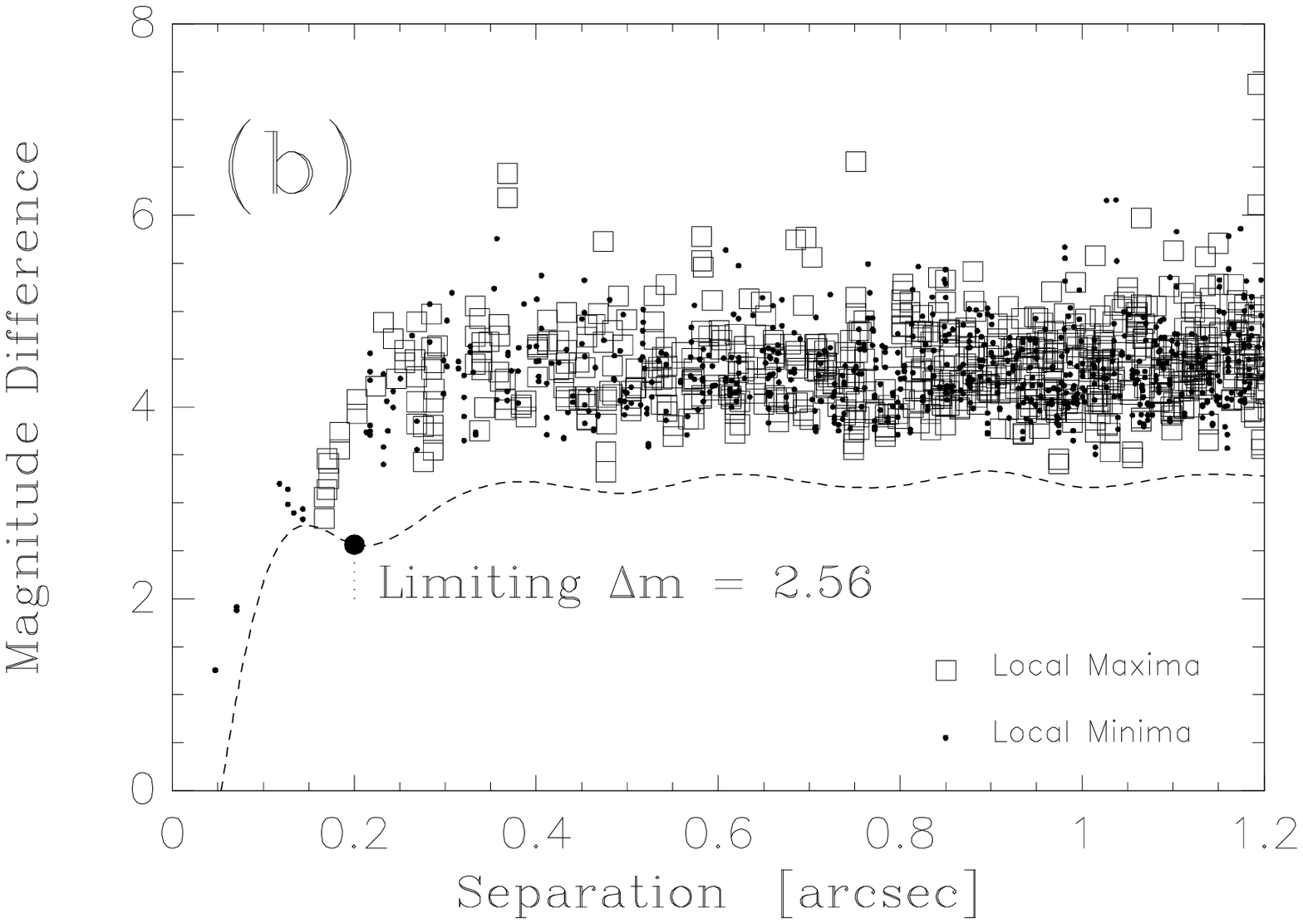}
}
\caption{
Limits on the relative magnitude of a resolved companion to HAT-P-67 as a function of angular separation based on speckle imaging observations from WIYN~3.5\,m/DSSI. The dotted lines denote the $5\sigma$ limits. The left panel shows the limits for the 692\,nm filter, the right shows limits for the 880\,nm filter. 
}
\label{fig:luckyimage}
\end{figure*}

The DSSI observations were gathered between the nights of UT 26 September 2015
and UT 3 October 2015. A dichroic beamsplitter was used to obtain
simultaneous imaging through 692\,nm and 880\,nm filters. Each
observation consisted of a sequence of 1000 40\,ms exposures read-out
on $128 \times 128$ pixel ($2\farcs8 \times 2\farcs8$) subframes,
which were reduced to reconstructed images following
\citet{horch:2011}. These images were searched for companions, with none detected. Based on this, the $5\sigma$ lower limits on the differential
magnitude between a putative companion and the primary star were
determined as a function of angular separation as described in
\citet{horch:2011}. Based on these observations we exclude neighbors with $\Delta m < 2.56$ at 692\,nm, or $\Delta m < 2.80$ at 880\,nm, down to a limiting separation of $0\farcs2$ (Figure~\ref{fig:luckyimage}).

\subsection{Blend analysis}
\label{sec:blend}

Blend scenarios are eliminated by the detection of the planetary Doppler tomographic transit signal. In the cases where an eclipsing binary blended with a foreground star is the cause of the transit signal, the Doppler tomographic shadow will be significantly diluted with respect to the photometric transit signal. 

The flux under the shadow of the planet, as a fraction of the total flux under the rotational broadening kernel, describes the area blocked by the planet. This directly corresponds to a `transit light curve' over the broadband of the TRES spectrum \citep[following][]{Zhou:2016}. We plot this Doppler tomographic light curve in Figure~\ref{fig:dopplertomography} (bottom). We also plot the model transit light curve as per the global best fit solution. The spectroscopic transit depth is consistent with that of the photometric transit depth, confirming the lack of any significant dilution by background stars. The elimination of blend scenarios and the mass upper limit determined from HIRES radial velocities validates \hatcurb{} as a planet. We can also place strict upper limits on any third light contamination from background stars by modelling the line broadening profiles from the LSD analysis. A high signal-to-noise broadening profile was derived by averaging the 32 TRES spectra obtained for \hatcur{}. By modeling this profile as two stars, we place an upper limit on the flux ratio of any potential companion to be $<0.004$, or within 6 magnitudes of the primary star, with the caveat that any potential blended companion exhibits no radial velocity variation.

\section{Discussion}
\label{sec:discussion}

We presented the discovery of \hatcurb{}, a hot-Saturn transiting an F-subgiant. \hatcurb{} has a radius of $\genevaplanetradius\,\rjup$, and a mass constrained by radial velocity measurements to be $M_p < 0.59\,M_J$ at $1\,\sigma$. Confirmation of the planetary nature of \hatcurb{} involved numerous high precision follow-up transit light curves, radial-velocity constraints on its mass, and two Doppler tomographic transits that eliminated potential blended eclipsing binary scenarios. 

\begin{figure*}[!ht]
\plottwo{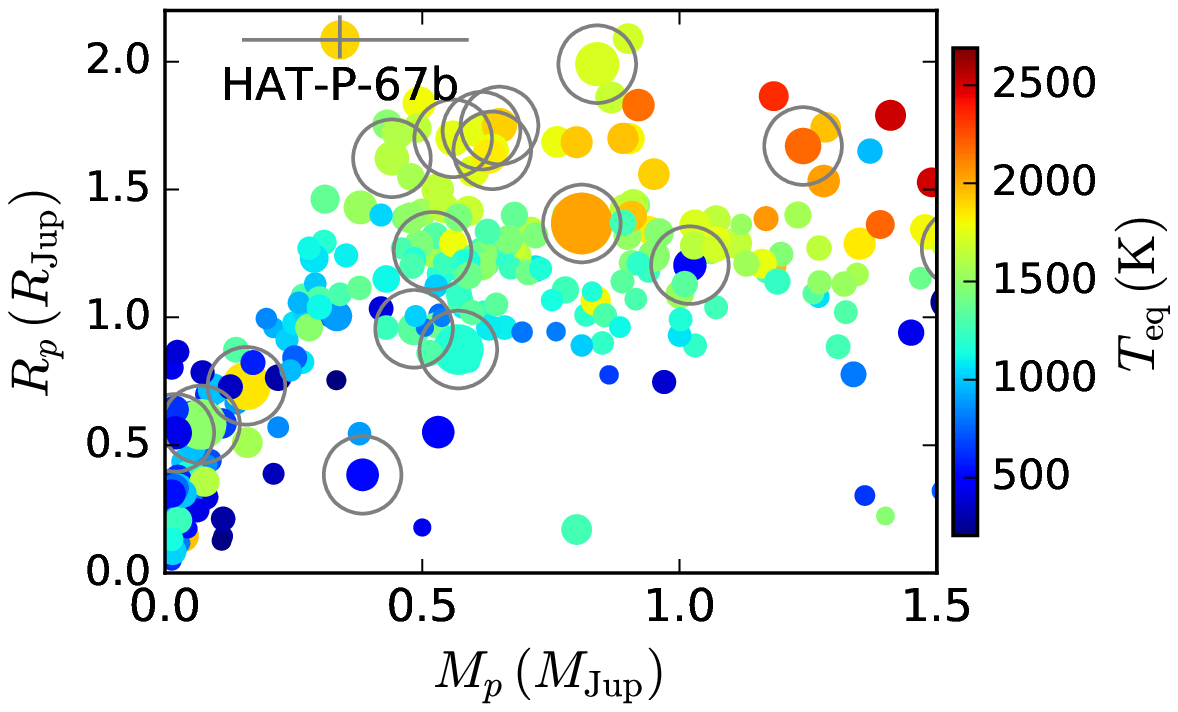}{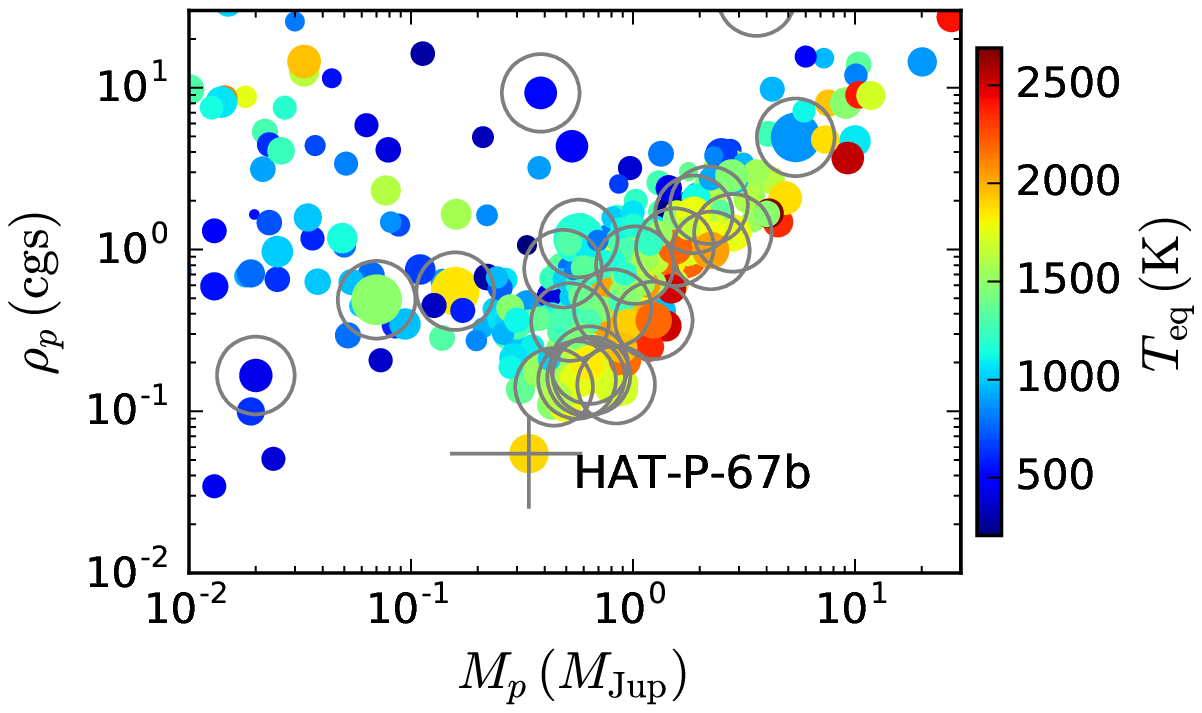}
\caption{
The mass--radius and mass--density distributions of known transiting exoplanets are plotted. The colors of the points represent the equilibrium temperatures of the planets, while their sizes are scaled to indicate the radii of the host stars. Planets that orbit evolved stars ($\logg < 4.0$) are marked by the open grey circles. \hatcurb{} is labelled, and its $1\sigma$ uncertainties are shown by the error bars. We note that it is one of the largest radius, lowest density planets found to date. 
\label{fig:planetpopulation}}
\end{figure*}

\begin{figure*}[!ht]
\plottwo{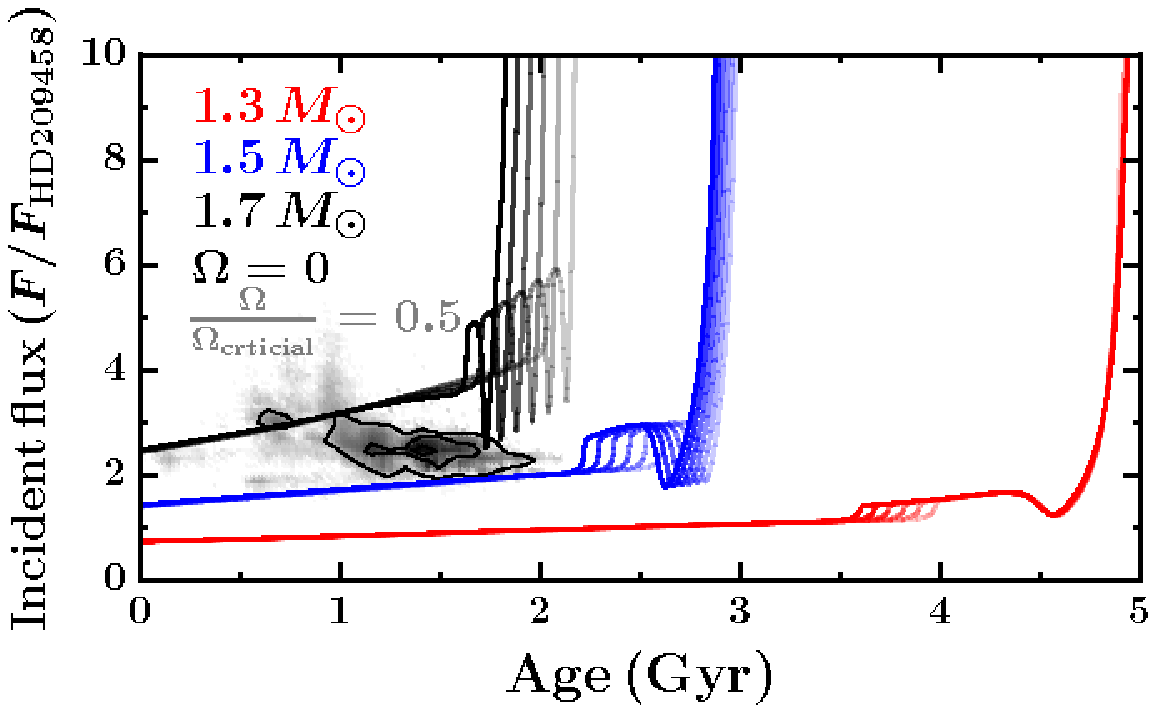}{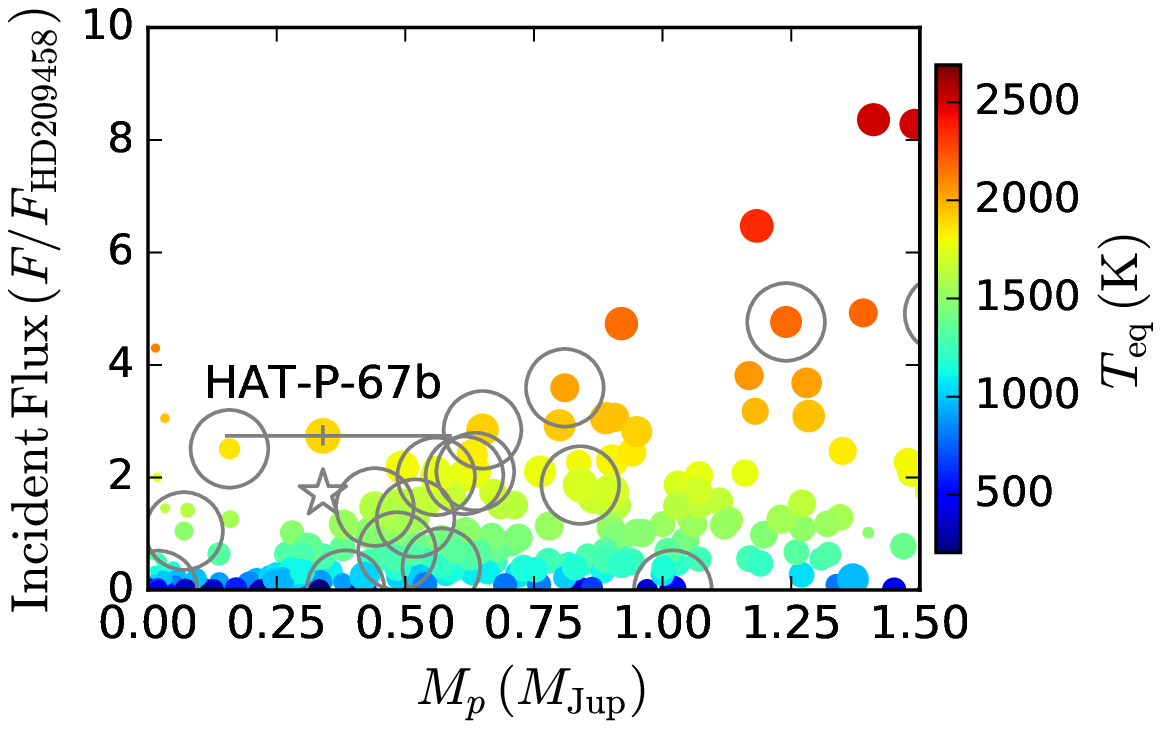}
\caption{
The incident flux received by \hatcurb{}. The \textbf{left} panel shows the changing incident flux of \hatcurb{}, calculated from the adopted Geneva isochrones. The line color and shading correspond with that shown in Figure~\ref{fig:iso}. The \textbf{right} panel shows the distribution of incident flux received by the transiting planet distribution, as a function of planet mass. The colors of individual points indicate their equilibrium temperatures, while the size of the points are scaled to the radii of the planets. The ZAMS incident flux of \hatcurb{} is marked by the grey star. Planets that orbit evolved stars $(\logg < 4.0)$ are marked by open grey circles. 
\label{fig:flux}}
\end{figure*}

The mass, radius, and densities of \hatcurb{} are plotted in Figure~\ref{fig:planetpopulation}, along with selected parts of the gas giant population. \hatcurb{} is one of the largest, and one of the lowest density planets known $(\rhopl = \genevaplrho \, \gcmc)$. A number of other inflated gas giants have been discovered around subgiants (KOI-680b \citealt{2015A&A...575A..71A}, EPIC 206247743b \citealt{2016arXiv160509180V}, KELT-8b \citealt{2015ApJ...810...30F}, KELT-11b \citealt{2016arXiv160701755P}, HAT-P-65b and HAT-P-66b \citealt{2016arXiv160902767H}) and giants (e.g. EPIC 211351816b \citealt{Grunblatt:2016}). One hypothesis is that these gas giants are re-inflated by the evolved host star \citep{2016ApJ...818....4L}. In this scenario, as the host star evolves off the main sequence, `warm Jupiters' are subjected to higher incident flux and stronger tidal heating (assuming a non-zero initial eccentricity). The heating reaches deep enough into the planetary interior to inflate the planet radius. \citet{2016arXiv160902767H}) found empirical evidence that the level of planet inflation is correlated with the fractional age of the host star, further supporting the idea of re-inflation. Figure~\ref{fig:flux} shows the evolution in the incident flux received by \hatcurb{} over its lifetime. Currently \hatcurb{} receives $\sim 2\times$ the incident-flux of a zero-age-main-sequence \hatcur{}, potentially inducing an inflation of the planetary radius.

Figure~\ref{fig:flux} also plots the incident flux received by the hot-Jupiter distribution against their planet masses. There is a paucity of low mass planets that receive high incident flux -- a sharp envelope that likely resulted from the evaporation of Saturn and Neptune mass planets in close-in orbits \citep[e.g.][]{2007A&A...461.1185L,2011A&A...529A.136E,2013ApJ...775..105O}. \hatcurb{} lies on the edge of the envelope -- unlike planets of similar masses that receive high incident irradiation, \hatcurb{} did not `boil-off', but survived to the present day. The high incident flux may also have halted contraction early on, leading to its current radius. Since there is a lack of inflated Saturn-mass planets in high incident flux environments, \hatcurb{} is an important point in the mass-radius-flux relationship. 

The low density and high irradiation of \hatcurb{} also results in a bloated atmosphere, with a large scale height of $\sim500$\,km (assuming an $H_2$ atmosphere), making the planet a good candidate for transmission spectroscopy follow-up studies.

\begin{figure*}[!ht]
\plottwo{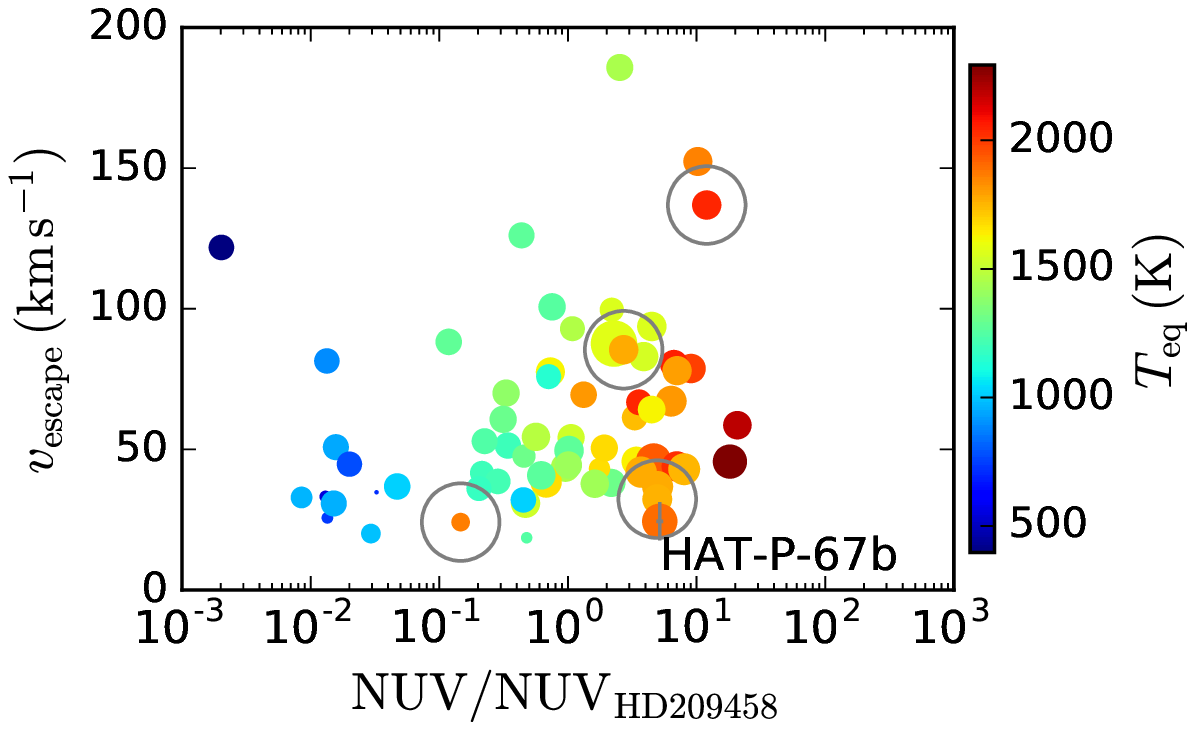}{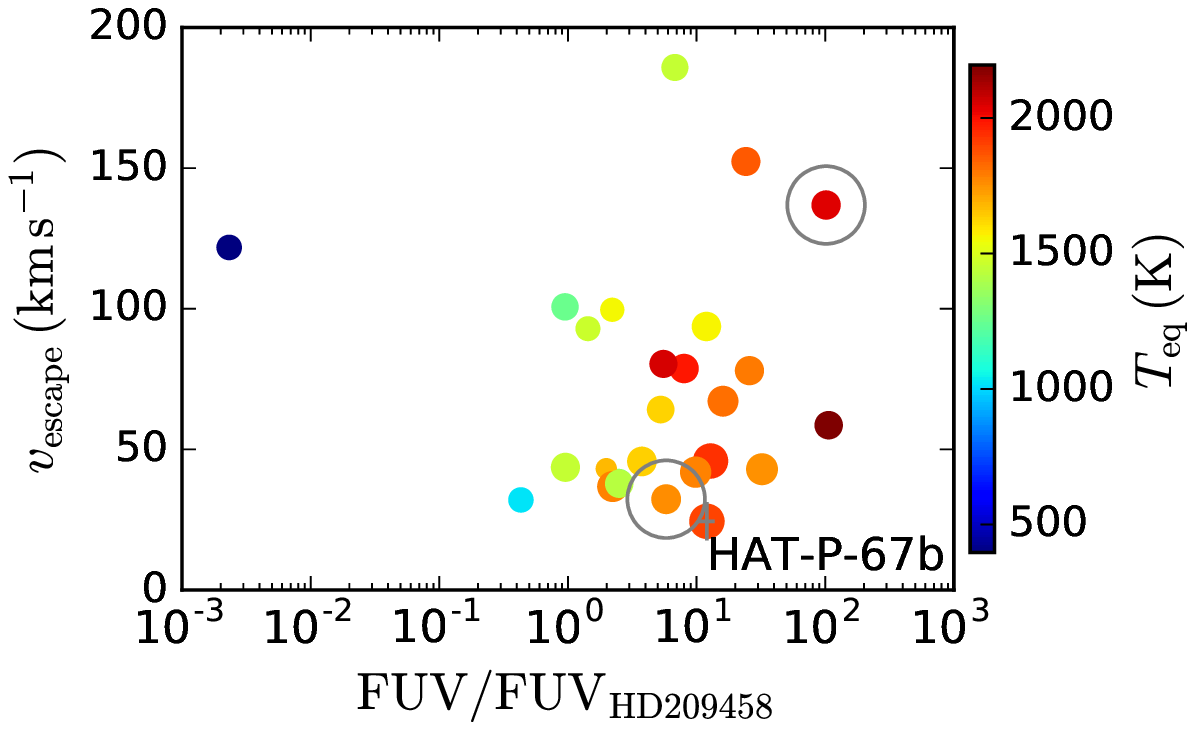}
\caption{
Mass loss is driven by UV and X-ray irradiation of the upper atmosphere of planets. \hatcurb{} potentially has one of the highest mass loss rates of known hot-Jupiters. We plot the NUV \textbf{left} and FUV \textbf{right} fluxes received by known transiting systems against their escape velocities. Only systems with GALEX UV fluxes \citep{Bianchi:2011} and GAIA parallaxes are plotted. As with Figure~\ref{fig:planetpopulation}, the sizes of the points indicate their planetary radii, while the colors represent the equilibrium temperatures. Planets orbiting evolved stars are marked by open grey circles. 
\label{fig:UV}}
\end{figure*}

In addition, X-ray and EUV-driven hydrodynamic escape play an especially important role in low density, low mass planets \citep[e.g.][]{2007A&A...461.1185L,2009ApJ...693...23M,2011A&A...529A.136E,2012MNRAS.425.2931O,2013ApJ...775..105O}. For hot-Jupiters, X-ray and EUV photoionizes the upper atmosphere, causing it to heat up and expand, resulting in escaping flows.  Atmospheric escape has been observed for HD 209458 b \citep{2003Natur.422..143V,2004ApJ...604L..69V} and HD 189733 b \citep{2010A&A...514A..72L}, where the Lyman-$\alpha$ radii of the planets are $\sim 10$ times larger than their optical radii, extending beyond the Roche sphere. Since the mass loss rate is largely proportional to the incident UV and X-ray flux received by the planets \citep[e.g.][]{2009ApJ...693...23M}, we checked for existing X-ray and UV measurements of \hatcur{}. While no EUV or X-ray flux measurements exist, \hatcur{} is identified as a source by GALEX, with flux measurements in the FUV (1344--1786\AA) and NUV (1771--2831\AA) bands. In Figure~\ref{fig:UV}, we compile all transiting planet systems with GALEX FUV and NUV measurements \citep{Bianchi:2011}, as well as GAIA parallaxes and updated system parameters from \citet{2016arXiv160904389S}. To examine the potential mass loss rate of \hatcurb{} in the context of existing systems, we plot the UV fluxes received by each planet (normalized to that received by HD 209458 b) against their escape velocities. \hatcurb{} receives 24 times the FUV, and 10 times the NUV flux of HD 209458 b, and has one of the lowest escape velocities of known transiting planets ($25\,\mathrm{km\,s}^{-1}$, compared to $43\,\mathrm{km\,s}^{-1}$ for HD 209458 b). As such, it should be an excellent target for Lyman-$\alpha$ transit observations to measure its extended hydrogen exosphere. 

\hatcur{} has the highest \vsini{} amongst all the evolved planet hosts ($\genevastarvsini\,\kms$), allowing us to spectroscopically measure its projected spin alignment angle. The only previous Holt-Rossiter-McLaughlin measurement of a planet around a subgiant was WASP-71b, \citep{2013A&A...552A.120S}, which was found to be in a well-aligned orbit. Two additional planet-hosting evolved stars have had their line-of-sight stellar inclination measured via asteroseismology. \citet{2015ApJ...803...49Q} found the super-Jupiter Kepler-432b to be spin-orbit aligned in a eccentric, 53-day period orbit around a red-giant. \citet{2013Sci...342..331H} found two co-planar planets residing in inclined orbits around their red-giant host Kepler-56.

As a host star expands over its post main-sequence evolution, star-planet tidal interaction should increase in strength, modifying the orbit of the planet. radial-velocity searches have found that eccentric planets are rarer around evolved hosts than around dwarfs \citep[e.g.][]{Jones:2014}. However, tidal interactions are likely weak during the lifetime of \hatcurb{}: the characteristic timescale for orbital decay is on the order of $\sim 10^{11}$ years \citep[adopting eq.2 of ][ assuming an effective stellar dissipation coefficient of $\sigma_\star = 10^{-8}$]{Hansen:2012}. Similarly, the tidal synchronization timescale -- the time taken to synchronize the stellar spin vector with the orbit normal vector, leading to spin-orbit alignment and spin-orbit synchronization, is $\sim 10^{13}$ years \citep[eq. 3 of ][]{Hansen:2012}, so the stellar spin is unlikely to have been modified over the lifetime of the system due to planet-star tidal interactions. The same cannot be said of WASP-71b, which orbits at a shorter period of 2.9 days, and a characteristic synchronization timescale of $\sim 10^{10}$ years, short enough that tides likely played a role in modifying the spin of the host star.

\hatcurb{} is part of a growing list of planets confirmed via Doppler tomography, including WASP-33b \citep{Collier:2010b,Johnson:2015}, Kepler-13b \citep{Johnson:2014}, KELT-7b \citep{Bieryla:2015,Zhou:2016} HAT-P-57b \citep{Hartman:2015}, KOI-12b \citep{Bourrier:2015}, KELT-17b \citep{2016arXiv160703512Z}, and XO-6b \citep{2016arXiv161202776C}. The routine success of the Doppler tomography technique yields exciting prospects to fill the paucity of transiting planets around early type stars.

\acknowledgements  

We thank the referee for their insightful contributions to this manuscript. HATNet operations have been
funded by NASA grants NNG04GN74G and NNX13AJ15G. Follow-up of HATNet
targets has been partially supported through NSF grant
AST-1108686. G.\'A.B., Z.C., and K.P.\ acknowledge partial support
from NASA grant NNX09AB29G. J.H.\ acknowledges support from NASA grant
NNX14AE87G. K.P.\ acknowledges support from NASA grant NNX13AQ62G. We
acknowledge partial support also from the {\em Kepler} Mission under
NASA Cooperative Agreement NCC2-1390 (D.W.L., PI). Data presented in this paper are
based on observations obtained at the HAT station at the Submillimeter
Array of SAO, and the HAT station at the Fred Lawrence Whipple
Observatory of SAO. This research has made use of the NASA Exoplanet
Archive, which is operated by the California Institute of Technology,
under contract with the National Aeronautics and Space Administration
under the Exoplanet Exploration Program. Data presented herein were
obtained at the WIYN Observatory from telescope time allocated to
NN-EXPLORE through the scientific partnership of the National
Aeronautics and Space Administration, the National Science Foundation,
and the National Optical Astronomy Observatory. This work was
supported by a NASA WIYN PI Data Award, administered by the NASA
Exoplanet Science Institute. We acknowledge of J.A. Johnson in supporting the Keck HIRES observations. The authors wish to recognize and
acknowledge the very significant cultural role and reverence that the
summit of Mauna Kea has always had within the indigenous Hawaiian
community. We are most fortunate to have the opportunity to conduct
observations from this mountain.

\bibliographystyle{apj}
\bibliography{hatbib}

\begin{thebibliography}{89}
\expandafter\ifx\csname natexlab\endcsname\relax\def\natexlab#1{#1}\fi

\bibitem[{{Almenara} {et~al.}(2015){Almenara}, {Damiani}, {Bouchy}, {Havel},
  {Bruno}, {H{\'e}brard}, {Diaz}, {Deleuil}, {Barros}, {Boisse}, {Bonomo},
  {Montagnier}, \& {Santerne}}]{2015A&A...575A..71A}
{Almenara}, J.~M., {Damiani}, C., {Bouchy}, F., {et~al.} 2015, \aap, 575, A71

\bibitem[{{Bakos} {et~al.}(2004){Bakos}, {Noyes}, {Kov{\'a}cs}, {Stanek},
  {Sasselov}, \& {Domsa}}]{Bakos:2004}
{Bakos}, G., {Noyes}, R.~W., {Kov{\'a}cs}, G., {et~al.} 2004, \pasp, 116, 266

\bibitem[{{Bakos} {et~al.}(2007){Bakos}, {Kov{\'a}cs}, {Torres}, {Fischer},
  {Latham}, {Noyes}, {Sasselov}, {Mazeh}, {Shporer}, {Butler}, {Stefanik},
  {Fern{\'a}ndez}, {Sozzetti}, {P{\'a}l}, {Johnson}, {Marcy}, {Winn}, {Sip{\H
  o}cz}, {L{\'a}z{\'a}r}, {Papp}, \& {S{\'a}ri}}]{2007ApJ...670..826B}
{Bakos}, G.~{\'A}., {Kov{\'a}cs}, G., {Torres}, G., {et~al.} 2007, \apj, 670,
  826

\bibitem[{{Bakos} {et~al.}(2010){Bakos}, {Torres}, {P{\'a}l}, {Hartman},
  {Kov{\'a}cs}, {Noyes}, {Latham}, {Sasselov}, {Sip{\H o}cz}, {Esquerdo},
  {Fischer}, {Johnson}, {Marcy}, {Butler}, {Isaacson}, {Howard}, {Vogt},
  {Kov{\'a}cs}, {Fernandez}, {Mo{\'o}r}, {Stefanik}, {L{\'a}z{\'a}r}, {Papp},
  \& {S{\'a}ri}}]{2010ApJ...710.1724B}
{Bakos}, G.~{\'A}., {Torres}, G., {P{\'a}l}, A., {et~al.} 2010, \apj, 710, 1724

\bibitem[{{B{\'e}ky} {et~al.}(2011){B{\'e}ky}, {Bakos}, {Hartman}, {Torres},
  {Latham}, {Jord{\'a}n}, {Arriagada}, {Bayliss}, {Kiss}, {Kov{\'a}cs},
  {Quinn}, {Marcy}, {Howard}, {Fischer}, {Johnson}, {Esquerdo}, {Noyes},
  {Buchhave}, {Sasselov}, {Stefanik}, {Perumpilly}, {L{\'a}z{\'a}r}, {Papp}, \&
  {S{\'a}ri}}]{2011ApJ...734..109B}
{B{\'e}ky}, B., {Bakos}, G.~{\'A}., {Hartman}, J., {et~al.} 2011, \apj, 734,
  109

\bibitem[{{Bianchi} {et~al.}(2011){Bianchi}, {Herald}, {Efremova}, {Girardi},
  {Zabot}, {Marigo}, {Conti}, \& {Shiao}}]{Bianchi:2011}
{Bianchi}, L., {Herald}, J., {Efremova}, B., {et~al.} 2011, \apss, 335, 161

\bibitem[{{Bieryla} {et~al.}(2015){Bieryla}, {Collins}, {Beatty}, {Eastman},
  {Siverd}, {Pepper}, {Gaudi}, {Stassun}, {Ca{\~n}as}, {Latham}, {Buchhave},
  {Sanchis-Ojeda}, {Winn}, {Jensen}, {Kielkopf}, {McLeod}, {Gregorio},
  {Col{\'o}n}, {Street}, {Ross}, {Penny}, {Mellon}, {Oberst}, {Fulton}, {Wang},
  {Berlind}, {Calkins}, {Esquerdo}, {DePoy}, {Gould}, {Marshall}, {Pogge},
  {Trueblood}, \& {Trueblood}}]{Bieryla:2015}
{Bieryla}, A., {Collins}, K., {Beatty}, T.~G., {et~al.} 2015, \aj, 150, 12

\bibitem[{{Bourrier} {et~al.}(2015){Bourrier}, {Lecavelier des Etangs},
  {H{\'e}brard}, {Santerne}, {Deleuil}, {Almenara}, {Barros}, {Boisse},
  {Bonomo}, {Bruno}, {Courcol}, {Diaz}, {Montagnier}, \&
  {Moutou}}]{Bourrier:2015}
{Bourrier}, V., {Lecavelier des Etangs}, A., {H{\'e}brard}, G., {et~al.} 2015,
  \aap, 579, A55

\bibitem[{{Bowler} {et~al.}(2010){Bowler}, {Johnson}, {Marcy}, {Henry}, {Peek},
  {Fischer}, {Clubb}, {Liu}, {Reffert}, {Schwab}, \&
  {Lowe}}]{2010ApJ...709..396B}
{Bowler}, B.~P., {Johnson}, J.~A., {Marcy}, G.~W., {et~al.} 2010, \apj, 709,
  396

\bibitem[{{Buchhave} {et~al.}(2010){Buchhave}, {Bakos}, {Hartman}, {Torres},
  {Kov{\'a}cs}, {Latham}, {Noyes}, {Esquerdo}, {Everett}, {Howard}, {Marcy},
  {Fischer}, {Johnson}, {Andersen}, {F{\H u}r{\'e}sz}, {Perumpilly},
  {Sasselov}, {Stefanik}, {B{\'e}ky}, {L{\'a}z{\'a}r}, {Papp}, \&
  {S{\'a}ri}}]{Buchave:2010}
{Buchhave}, L.~A., {Bakos}, G.~{\'A}., {Hartman}, J.~D., {et~al.} 2010, \apj,
  720, 1118

\bibitem[{{Buchhave} {et~al.}(2012){Buchhave}, {Latham}, {Johansen},
  {Bizzarro}, {Torres}, {Rowe}, {Batalha}, {Borucki}, {Brugamyer}, {Caldwell},
  {Bryson}, {Ciardi}, {Cochran}, {Endl}, {Esquerdo}, {Ford}, {Geary},
  {Gilliland}, {Hansen}, {Isaacson}, {Laird}, {Lucas}, {Marcy}, {Morse},
  {Robertson}, {Shporer}, {Stefanik}, {Still}, \& {Quinn}}]{Buchave:2012}
{Buchhave}, L.~A., {Latham}, D.~W., {Johansen}, A., {et~al.} 2012, \nat, 486,
  375

\bibitem[{{Butler} {et~al.}(1996){Butler}, {Marcy}, {Williams}, {McCarthy},
  {Dosanjh}, \& {Vogt}}]{1996PASP..108..500B}
{Butler}, R.~P., {Marcy}, G.~W., {Williams}, E., {et~al.} 1996, \pasp, 108, 500

\bibitem[{{Casertano} {et~al.}(2016){Casertano}, {Riess}, {Bucciarelli}, \&
  {Lattanzi}}]{2016arXiv160905175C}
{Casertano}, S., {Riess}, A.~G., {Bucciarelli}, B., \& {Lattanzi}, M.~G. 2016,
  ArXiv e-prints, 1609.05175

\bibitem[{{Castelli} \& {Kurucz}(2004)}]{Castelli:2004}
{Castelli}, F., \& {Kurucz}, R.~L. 2004, ArXiv Astrophysics e-prints

\bibitem[{{Ciceri} {et~al.}(2015){Ciceri}, {Lillo-Box}, {Southworth},
  {Mancini}, {Henning}, \& {Barrado}}]{2015A&A...573L...5C}
{Ciceri}, S., {Lillo-Box}, J., {Southworth}, J., {et~al.} 2015, \aap, 573, L5

\bibitem[{{Claret} \& {Bloemen}(2011)}]{Claret:2011}
{Claret}, A., \& {Bloemen}, S. 2011, \aap, 529, A75

\bibitem[{{Collier Cameron} {et~al.}(2010{\natexlab{a}}){Collier Cameron},
  {Bruce}, {Miller}, {Triaud}, \& {Queloz}}]{Collier:2010a}
{Collier Cameron}, A., {Bruce}, V.~A., {Miller}, G.~R.~M., {Triaud},
  A.~H.~M.~J., \& {Queloz}, D. 2010{\natexlab{a}}, \mnras, 403, 151

\bibitem[{{Collier Cameron} {et~al.}(2010{\natexlab{b}}){Collier Cameron},
  {Guenther}, {Smalley}, {McDonald}, {Hebb}, {Andersen}, {Augusteijn},
  {Barros}, {Brown}, {Cochran}, {Endl}, {Fossey}, {Hartmann}, {Maxted},
  {Pollacco}, {Skillen}, {Telting}, {Waldmann}, \& {West}}]{Collier:2010b}
{Collier Cameron}, A., {Guenther}, E., {Smalley}, B., {et~al.}
  2010{\natexlab{b}}, \mnras, 407, 507

\bibitem[{{Crouzet} {et~al.}(2016){Crouzet}, {McCullough}, {Long}, {Montanes
  Rodriguez}, {Lecavelier des Etangs}, {Ribas}, {Bourrier}, {H{\'e}brard},
  {Vilardell}, {Deleuil}, {Herrero}, {Garcia-Melendo}, {Akhenak}, {Foote},
  {Gary}, {Benni}, {Guillot}, {Conjat}, {M{\'e}karnia}, {Garlitz}, {Burke},
  {Courcol}, \& {Demangeon}}]{2016arXiv161202776C}
{Crouzet}, N., {McCullough}, P.~R., {Long}, D., {et~al.} 2016, ArXiv e-prints,
  1612.02776

\bibitem[{{Dawson} \& {Johnson}(2012)}]{2012ApJ...756..122D}
{Dawson}, R.~I., \& {Johnson}, J.~A. 2012, \apj, 756, 122

\bibitem[{{Dobbs-Dixon} {et~al.}(2004){Dobbs-Dixon}, {Lin}, \&
  {Mardling}}]{2004ApJ...610..464D}
{Dobbs-Dixon}, I., {Lin}, D.~N.~C., \& {Mardling}, R.~A. 2004, \apj, 610, 464

\bibitem[{{Donati} {et~al.}(1997){Donati}, {Semel}, {Carter}, {Rees}, \&
  {Collier Cameron}}]{Donati:1997}
{Donati}, J.-F., {Semel}, M., {Carter}, B.~D., {Rees}, D.~E., \& {Collier
  Cameron}, A. 1997, \mnras, 291, 658

\bibitem[{{Dotter} {et~al.}(2008){Dotter}, {Chaboyer}, {Jevremovi{\'c}},
  {Kostov}, {Baron}, \& {Ferguson}}]{2008ApJS..178...89D}
{Dotter}, A., {Chaboyer}, B., {Jevremovi{\'c}}, D., {et~al.} 2008, \apjs, 178,
  89

\bibitem[{{Ehrenreich} \& {D{\'e}sert}(2011)}]{2011A&A...529A.136E}
{Ehrenreich}, D., \& {D{\'e}sert}, J.-M. 2011, \aap, 529, A136

\bibitem[{{Ekstr{\"o}m} {et~al.}(2012){Ekstr{\"o}m}, {Georgy}, {Eggenberger},
  {Meynet}, {Mowlavi}, {Wyttenbach}, {Granada}, {Decressin}, {Hirschi},
  {Frischknecht}, {Charbonnel}, \& {Maeder}}]{2012A&A...537A.146E}
{Ekstr{\"o}m}, S., {Georgy}, C., {Eggenberger}, P., {et~al.} 2012, \aap, 537,
  A146

\bibitem[{{Enoch} {et~al.}(2012){Enoch}, {Collier Cameron}, \&
  {Horne}}]{2012A&A...540A..99E}
{Enoch}, B., {Collier Cameron}, A., \& {Horne}, K. 2012, \aap, 540, A99

\bibitem[{{Espinoza} {et~al.}(2016){Espinoza}, {Bayliss}, {Hartman}, {Bakos},
  {Jord{\'a}n}, {Zhou}, {Mancini}, {Brahm}, {Ciceri}, {Bhatti}, {Csubry},
  {Rabus}, {Penev}, {Bento}, {de Val-Borro}, {Henning}, {Schmidt}, {Suc},
  {Wright}, {Tinney}, {Tan}, \& {Noyes}}]{2016AJ....152..108E}
{Espinoza}, N., {Bayliss}, D., {Hartman}, J.~D., {et~al.} 2016, \aj, 152, 108

\bibitem[{{{F{\H u}r{\'e}sz}}(2008)}]{Furesz:2008}
{{F{\H u}r{\'e}sz}}, G. 2008, PhD thesis, Univ. of Szeged, Hungary

\bibitem[{{Foreman-Mackey} {et~al.}(2013){Foreman-Mackey}, {Hogg}, {Lang}, \&
  {Goodman}}]{ForemanMackey:2012}
{Foreman-Mackey}, D., {Hogg}, D.~W., {Lang}, D., \& {Goodman}, J. 2013, \pasp,
  125, 306

\bibitem[{{Freed} {et~al.}(2004){Freed}, {Hinz}, {Meyer}, {Milton}, \&
  {Lloyd-Hart}}]{freed:2004}
{Freed}, M., {Hinz}, P.~M., {Meyer}, M.~R., {Milton}, N.~M., \& {Lloyd-Hart},
  M. 2004, in Society of Photo-Optical Instrumentation Engineers (SPIE)
  Conference Series, Vol. 5492, Ground-based Instrumentation for Astronomy, ed.
  A.~F.~M. {Moorwood} \& M.~{Iye}, 1561--1571

\bibitem[{{Fulton} {et~al.}(2015){Fulton}, {Collins}, {Gaudi}, {Stassun},
  {Pepper}, {Beatty}, {Siverd}, {Penev}, {Howard}, {Baranec}, {Corfini},
  {Eastman}, {Gregorio}, {Law}, {Lund}, {Oberst}, {Penny}, {Riddle},
  {Rodriguez}, {Stevens}, {Zambelli}, {Ziegler}, {Bieryla}, {D'Ago}, {DePoy},
  {Jensen}, {Kielkopf}, {Latham}, {Manner}, {Marshall}, {McLeod}, \&
  {Reed}}]{2015ApJ...810...30F}
{Fulton}, B.~J., {Collins}, K.~A., {Gaudi}, B.~S., {et~al.} 2015, \apj, 810, 30

\bibitem[{{Gray} \& {Corbally}(1994)}]{Gray:1994}
{Gray}, R.~O., \& {Corbally}, C.~J. 1994, \aj, 107, 742

\bibitem[{{Grunblatt} {et~al.}(2016){Grunblatt}, {Huber}, {Gaidos}, {Lopez},
  {Fulton}, {Fortney}, {Howard}, {Sinukoff}, {Mann}, \&
  {Isaacson}}]{Grunblatt:2016}
{Grunblatt}, S.~K., {Huber}, D., {Gaidos}, E.~J., {et~al.} 2016, ArXiv
  e-prints, 1606.05818

\bibitem[{{Hansen}(2012)}]{Hansen:2012}
{Hansen}, B.~M.~S. 2012, \apj, 757, 6

\bibitem[{{Hansen} \& {Barman}(2007)}]{2007ApJ...671..861H}
{Hansen}, B.~M.~S., \& {Barman}, T. 2007, \apj, 671, 861

\bibitem[{{Hartman} {et~al.}(2011){Hartman}, {Bakos}, {Torres}, {Latham},
  {Kov{\'a}cs}, {B{\'e}ky}, {Quinn}, {Mazeh}, {Shporer}, {Marcy}, {Howard},
  {Fischer}, {Johnson}, {Esquerdo}, {Noyes}, {Sasselov}, {Stefanik},
  {Fernandez}, {Szklen{\'a}r}, {L{\'a}z{\'a}r}, {Papp}, \&
  {S{\'a}ri}}]{2011ApJ...742...59H}
{Hartman}, J.~D., {Bakos}, G.~{\'A}., {Torres}, G., {et~al.} 2011, \apj, 742,
  59

\bibitem[{{Hartman} {et~al.}(2015){Hartman}, {Bakos}, {Buchhave}, {Torres},
  {Latham}, {Kov{\'a}cs}, {Bhatti}, {Csubry}, {de Val-Borro}, {Penev}, {Huang},
  {B{\'e}ky}, {Bieryla}, {Quinn}, {Howard}, {Marcy}, {Johnson}, {Isaacson},
  {Fischer}, {Noyes}, {Falco}, {Esquerdo}, {Knox}, {Hinz}, {L{\'a}z{\'a}r},
  {Papp}, \& {S{\'a}ri}}]{Hartman:2015}
{Hartman}, J.~D., {Bakos}, G.~{\'A}., {Buchhave}, L.~A., {et~al.} 2015, \aj,
  150, 197

\bibitem[{{Hartman} {et~al.}(2016){Hartman}, {Bakos}, {Bhatti}, {Penev},
  {Bieryla}, {Latham}, {Kov{\'a}cs}, {Torres}, {Csubry}, {de Val-Borro},
  {Buchhave}, {Kov{\'a}cs}, {Quinn}, {Howard}, {Isaacson}, {Fulton}, {Everett},
  {Esquerdo}, {B{\'e}ky}, {Szklenar}, {Falco}, {Santerne}, {Boisse},
  {H{\'e}brard}, {Burrows}, {L{\'a}z{\'a}r}, {Papp}, \&
  {S{\'a}ri}}]{2016arXiv160902767H}
{Hartman}, J.~D., {Bakos}, G.~{\'A}., {Bhatti}, W., {et~al.} 2016, ArXiv
  e-prints, 1609.02767

\bibitem[{{Holt}(1893)}]{1893AstAp..12..646H}
{Holt}, J.~R. 1893, Astronomy and Astro-Physics (formerly The Sidereal
  Messenger), 12, 646

\bibitem[{{Horch} {et~al.}(2012){Horch}, {Bahi}, {Gaulin}, {Howell}, {Sherry},
  {Baena Gall{\'e}}, \& {van Altena}}]{horch:2012}
{Horch}, E.~P., {Bahi}, L.~A.~P., {Gaulin}, J.~R., {et~al.} 2012, \aj, 143, 10

\bibitem[{{Horch} {et~al.}(2011){Horch}, {van Altena}, {Howell}, {Sherry}, \&
  {Ciardi}}]{horch:2011}
{Horch}, E.~P., {van Altena}, W.~F., {Howell}, S.~B., {Sherry}, W.~H., \&
  {Ciardi}, D.~R. 2011, \aj, 141, 180

\bibitem[{{Howell} {et~al.}(2011){Howell}, {Everett}, {Sherry}, {Horch}, \&
  {Ciardi}}]{howell:2011}
{Howell}, S.~B., {Everett}, M.~E., {Sherry}, W., {Horch}, E., \& {Ciardi},
  D.~R. 2011, \aj, 142, 19

\bibitem[{{Huber} {et~al.}(2013){Huber}, {Carter}, {Barbieri}, {Miglio},
  {Deck}, {Fabrycky}, {Montet}, {Buchhave}, {Chaplin}, {Hekker},
  {Montalb{\'a}n}, {Sanchis-Ojeda}, {Basu}, {Bedding}, {Campante},
  {Christensen-Dalsgaard}, {Elsworth}, {Stello}, {Arentoft}, {Ford},
  {Gilliland}, {Handberg}, {Howard}, {Isaacson}, {Johnson}, {Karoff},
  {Kawaler}, {Kjeldsen}, {Latham}, {Lund}, {Lundkvist}, {Marcy}, {Metcalfe},
  {Silva Aguirre}, \& {Winn}}]{2013Sci...342..331H}
{Huber}, D., {Carter}, J.~A., {Barbieri}, M., {et~al.} 2013, Science, 342, 331

\bibitem[{{Jao} {et~al.}(2016){Jao}, {Henry}, {Riedel}, {Winters}, {Slatten},
  \& {Gies}}]{2016ApJ...832L..18J}
{Jao}, W.-C., {Henry}, T.~J., {Riedel}, A.~R., {et~al.} 2016, \apjl, 832, L18

\bibitem[{{Johnson} {et~al.}(2010){Johnson}, {Aller}, {Howard}, \&
  {Crepp}}]{Johnson:2010}
{Johnson}, J.~A., {Aller}, K.~M., {Howard}, A.~W., \& {Crepp}, J.~R. 2010,
  \pasp, 122, 905

\bibitem[{{Johnson} {et~al.}(2007){Johnson}, {Butler}, {Marcy}, {Fischer},
  {Vogt}, {Wright}, \& {Peek}}]{Johnson:2007}
{Johnson}, J.~A., {Butler}, R.~P., {Marcy}, G.~W., {et~al.} 2007, \apj, 670,
  833

\bibitem[{{Johnson} {et~al.}(2014){Johnson}, {Cochran}, {Albrecht},
  {Dodson-Robinson}, {Winn}, \& {Gullikson}}]{Johnson:2014}
{Johnson}, M.~C., {Cochran}, W.~D., {Albrecht}, S., {et~al.} 2014, \apj, 790,
  30

\bibitem[{{Johnson} {et~al.}(2015){Johnson}, {Cochran}, {Collier Cameron}, \&
  {Bayliss}}]{Johnson:2015}
{Johnson}, M.~C., {Cochran}, W.~D., {Collier Cameron}, A., \& {Bayliss}, D.
  2015, \apjl, 810, L23

\bibitem[{{Jones} {et~al.}(2014){Jones}, {Jenkins}, {Bluhm}, {Rojo}, \&
  {Melo}}]{Jones:2014}
{Jones}, M.~I., {Jenkins}, J.~S., {Bluhm}, P., {Rojo}, P., \& {Melo}, C.~H.~F.
  2014, \aap, 566, A113

\bibitem[{{Kipping}(2008)}]{2008MNRAS.389.1383K}
{Kipping}, D.~M. 2008, \mnras, 389, 1383

\bibitem[{{Kipping}(2013)}]{2013MNRAS.434L..51K}
---. 2013, \mnras, 434, L51

\bibitem[{{Kov{\'a}cs} {et~al.}(2005){Kov{\'a}cs}, {Bakos}, \&
  {Noyes}}]{Kovacs:2005}
{Kov{\'a}cs}, G., {Bakos}, G., \& {Noyes}, R.~W. 2005, \mnras, 356, 557

\bibitem[{{Kov{\'a}cs} {et~al.}(2002){Kov{\'a}cs}, {Zucker}, \&
  {Mazeh}}]{Kovacs:2002}
{Kov{\'a}cs}, G., {Zucker}, S., \& {Mazeh}, T. 2002, \aap, 391, 369

\bibitem[{{Lecavelier Des Etangs}(2007)}]{2007A&A...461.1185L}
{Lecavelier Des Etangs}, A. 2007, \aap, 461, 1185

\bibitem[{{Lecavelier des Etangs} {et~al.}(2004){Lecavelier des Etangs},
  {Vidal-Madjar}, {McConnell}, \& {H{\'e}brard}}]{2004A&A...418L...1L}
{Lecavelier des Etangs}, A., {Vidal-Madjar}, A., {McConnell}, J.~C., \&
  {H{\'e}brard}, G. 2004, \aap, 418, L1

\bibitem[{{Lecavelier Des Etangs} {et~al.}(2010){Lecavelier Des Etangs},
  {Ehrenreich}, {Vidal-Madjar}, {Ballester}, {D{\'e}sert}, {Ferlet},
  {H{\'e}brard}, {Sing}, {Tchakoumegni}, \& {Udry}}]{2010A&A...514A..72L}
{Lecavelier Des Etangs}, A., {Ehrenreich}, D., {Vidal-Madjar}, A., {et~al.}
  2010, \aap, 514, A72

\bibitem[{{Lillo-Box} {et~al.}(2014){Lillo-Box}, {Barrado}, {Moya},
  {Montesinos}, {Montalb{\'a}n}, {Bayo}, {Barbieri}, {R{\'e}gulo}, {Mancini},
  {Bouy}, \& {Henning}}]{2014A&A...562A.109L}
{Lillo-Box}, J., {Barrado}, D., {Moya}, A., {et~al.} 2014, \aap, 562, A109

\bibitem[{{Lindegren} {et~al.}(2016){Lindegren}, {Lammers}, {Bastian},
  {Hern{\'a}ndez}, {Klioner}, {Hobbs}, {Bombrun}, {Michalik}, {Ramos-Lerate},
  {Butkevich}, {Comoretto}, {Joliet}, {Holl}, {Hutton}, {Parsons},
  {Steidelm{\"u}ller}, {Abbas}, {Altmann}, {Andrei}, {Anton}, {Bach},
  {Barache}, {Becciani}, {Berthier}, {Bianchi}, {Biermann}, {Bouquillon},
  {Bourda}, {Br{\"u}semeister}, {Bucciarelli}, {Busonero}, {Carlucci},
  {Casta{\~n}eda}, {Charlot}, {Clotet}, {Crosta}, {Davidson}, {de Felice},
  {Drimmel}, {Fabricius}, {Fienga}, {Figueras}, {Fraile}, {Gai}, {Garralda},
  {Geyer}, {Gonz{\'a}lez-Vidal}, {Guerra}, {Hambly}, {Hauser}, {Jordan},
  {Lattanzi}, {Lenhardt}, {Liao}, {L{\"o}ffler}, {McMillan}, {Mignard}, {Mora},
  {Morbidelli}, {Portell}, {Riva}, {Sarasso}, {Serraller}, {Siddiqui}, {Smart},
  {Spagna}, {Stampa}, {Steele}, {Taris}, {Torra}, {van Reeven}, {Vecchiato},
  {Zschocke}, {de Bruijne}, {Gracia}, {Raison}, {Lister}, {Marchant},
  {Messineo}, {Soffel}, {Osorio}, {de Torres}, \&
  {O'Mullane}}]{2016arXiv160904303L}
{Lindegren}, L., {Lammers}, U., {Bastian}, U., {et~al.} 2016, ArXiv e-prints,
  1609.04303

\bibitem[{{Lopez} \& {Fortney}(2016)}]{2016ApJ...818....4L}
{Lopez}, E.~D., \& {Fortney}, J.~J. 2016, \apj, 818, 4

\bibitem[{{Lovis} \& {Mayor}(2007)}]{2007A&A...472..657L}
{Lovis}, C., \& {Mayor}, M. 2007, \aap, 472, 657

\bibitem[{{Mandel} \& {Agol}(2002)}]{2002ApJ...580L.171M}
{Mandel}, K., \& {Agol}, E. 2002, \apjl, 580, L171

\bibitem[{{McLaughlin}(1924)}]{McLaughlin:1924}
{McLaughlin}, D.~B. 1924, \apj, 60, 22

\bibitem[{{Meynet} \& {Maeder}(2000)}]{2000A&A...361..101M}
{Meynet}, G., \& {Maeder}, A. 2000, \aap, 361, 101

\bibitem[{{Murray-Clay} {et~al.}(2009){Murray-Clay}, {Chiang}, \&
  {Murray}}]{2009ApJ...693...23M}
{Murray-Clay}, R.~A., {Chiang}, E.~I., \& {Murray}, N. 2009, \apj, 693, 23

\bibitem[{{Muzerolle} {et~al.}(2003){Muzerolle}, {Hillenbrand}, {Calvet},
  {Brice{\~n}o}, \& {Hartmann}}]{Muzerolle:2003}
{Muzerolle}, J., {Hillenbrand}, L., {Calvet}, N., {Brice{\~n}o}, C., \&
  {Hartmann}, L. 2003, \apj, 592, 266

\bibitem[{{Natta} {et~al.}(2006){Natta}, {Testi}, \& {Randich}}]{Natta:2006}
{Natta}, A., {Testi}, L., \& {Randich}, S. 2006, \aap, 452, 245

\bibitem[{{Owen} \& {Jackson}(2012)}]{2012MNRAS.425.2931O}
{Owen}, J.~E., \& {Jackson}, A.~P. 2012, \mnras, 425, 2931

\bibitem[{{Owen} \& {Wu}(2013)}]{2013ApJ...775..105O}
{Owen}, J.~E., \& {Wu}, Y. 2013, \apj, 775, 105

\bibitem[{{Pepper} {et~al.}(2016){Pepper}, {Rodriguez}, {Collins}, {Johnson},
  {Fulton}, {Howard}, {Beatty}, {Stassun}, {Isaacson}, {Col{\'o}n}, {Lund},
  {Kuhn}, {Siverd}, {Gaudi}, {Tan}, {Curtis}, {Stockdale}, {Mawet}, {Bottom},
  {James}, {Zhou}, {Bayliss}, {Cargile}, {Bieryla}, {Penev}, {Latham},
  {Labadie-Bartz}, {Kielkopf}, {Eastman}, {Oberst}, {Jensen}, {Nelson},
  {Sliski}, {Wittenmyer}, {McCrady}, {Wright}, \&
  {Relles}}]{2016arXiv160701755P}
{Pepper}, J., {Rodriguez}, J.~E., {Collins}, K.~A., {et~al.} 2016, ArXiv
  e-prints, 1607.01755

\bibitem[{{Quinn} {et~al.}(2015){Quinn}, {White}, {Latham}, {Chaplin},
  {Handberg}, {Huber}, {Kipping}, {Payne}, {Jiang}, {Silva Aguirre}, {Stello},
  {Sliski}, {Ciardi}, {Buchhave}, {Bedding}, {Davies}, {Hekker}, {Kjeldsen},
  {Kuszlewicz}, {Everett}, {Howell}, {Basu}, {Campante},
  {Christensen-Dalsgaard}, {Elsworth}, {Karoff}, {Kawaler}, {Lund},
  {Lundkvist}, {Esquerdo}, {Calkins}, \& {Berlind}}]{2015ApJ...803...49Q}
{Quinn}, S.~N., {White}, T.~R., {Latham}, D.~W., {et~al.} 2015, \apj, 803, 49

\bibitem[{{Rossiter}(1924)}]{Rossiter:1924}
{Rossiter}, R.~A. 1924, \apj, 60, 15

\bibitem[{{Seager} \& {Mall{\'e}n-Ornelas}(2003)}]{Seager:2003}
{Seager}, S., \& {Mall{\'e}n-Ornelas}, G. 2003, in Astronomical Society of the
  Pacific Conference Series, Vol. 294, Scientific Frontiers in Research on
  Extrasolar Planets, ed. D.~{Deming} \& S.~{Seager}, 419--422

\bibitem[{{Shporer} {et~al.}(2011){Shporer}, {Jenkins}, {Rowe}, {Sanderfer},
  {Seader}, {Smith}, {Still}, {Thompson}, {Twicken}, \& {Welsh}}]{Shporer:2011}
{Shporer}, A., {Jenkins}, J.~M., {Rowe}, J.~F., {et~al.} 2011, \aj, 142, 195

\bibitem[{{Silva Aguirre} {et~al.}(2016){Silva Aguirre}, {Lund}, {Antia},
  {Ball}, {Basu}, {Christensen-Dalsgaard}, {Lebreton}, {Reese}, {Verma},
  {Casagrande}, {Justesen}, {Mosumgaard}, {Chaplin}, {Bedding}, {Davies},
  {Handberg}, {Houdek}, {Huber}, {Kjeldsen}, {Latham}, {White}, {Coelho},
  {Miglio}, \& {Rendle}}]{2016arXiv161108776S}
{Silva Aguirre}, V., {Lund}, M.~N., {Antia}, H.~M., {et~al.} 2016, ArXiv
  e-prints, 1611.08776

\bibitem[{{Skrutskie} {et~al.}(2006){Skrutskie}, {Cutri}, {Stiening},
  {Weinberg}, {Schneider}, {Carpenter}, {Beichman}, {Capps}, {Chester},
  {Elias}, {Huchra}, {Liebert}, {Lonsdale}, {Monet}, {Price}, {Seitzer},
  {Jarrett}, {Kirkpatrick}, {Gizis}, {Howard}, {Evans}, {Fowler}, {Fullmer},
  {Hurt}, {Light}, {Kopan}, {Marsh}, {McCallon}, {Tam}, {Van Dyk}, \&
  {Wheelock}}]{Skrutskie:2006}
{Skrutskie}, M.~F., {Cutri}, R.~M., {Stiening}, R., {et~al.} 2006, \aj, 131,
  1163

\bibitem[{{Smith} {et~al.}(2013){Smith}, {Anderson}, {Bouchy}, {Collier
  Cameron}, {Doyle}, {Fumel}, {Gillon}, {H{\'e}brard}, {Hellier}, {Jehin},
  {Lendl}, {Maxted}, {Moutou}, {Pepe}, {Pollacco}, {Queloz}, {Santerne},
  {Segransan}, {Smalley}, {Southworth}, {Triaud}, {Udry}, \&
  {West}}]{2013A&A...552A.120S}
{Smith}, A.~M.~S., {Anderson}, D.~R., {Bouchy}, F., {et~al.} 2013, \aap, 552,
  A120

\bibitem[{{Sozzetti} {et~al.}(2007){Sozzetti}, {Torres}, {Charbonneau},
  {Latham}, {Holman}, {Winn}, {Laird}, \& {O'Donovan}}]{Sozzetti:2007}
{Sozzetti}, A., {Torres}, G., {Charbonneau}, D., {et~al.} 2007, \apj, 664, 1190

\bibitem[{{Stassun} {et~al.}(2016){Stassun}, {Collins}, \&
  {Gaudi}}]{2016arXiv160904389S}
{Stassun}, K.~G., {Collins}, K.~A., \& {Gaudi}, B.~S. 2016, ArXiv e-prints,
  1609.04389

\bibitem[{{Stassun} \& {Torres}(2016)}]{2016ApJ...831L...6S}
{Stassun}, K.~G., \& {Torres}, G. 2016, \apjl, 831, L6

\bibitem[{{Szab{\'o}} {et~al.}(2011){Szab{\'o}}, {Szab{\'o}}, {Benk{\H o}},
  {Lehmann}, {Mez{\H o}}, {Simon}, {K{\H o}v{\'a}ri}, {Hodos{\'a}n},
  {Reg{\'a}ly}, \& {Kiss}}]{Szabo:2011}
{Szab{\'o}}, G.~M., {Szab{\'o}}, R., {Benk{\H o}}, J.~M., {et~al.} 2011, \apjl,
  736, L4

\bibitem[{{Telting} {et~al.}(2014){Telting}, {Avila}, {Buchhave}, {Frandsen},
  {Gandolfi}, {Lindberg}, {Stempels}, {Prins}, \& {NOT
  staff}}]{2014AN....335...41T}
{Telting}, J.~H., {Avila}, G., {Buchhave}, L., {et~al.} 2014, Astronomische
  Nachrichten, 335, 41

\bibitem[{{Torres} {et~al.}(2007){Torres}, {Bakos}, {Kov{\'a}cs}, {Latham},
  {Fern{\'a}ndez}, {Noyes}, {Esquerdo}, {Sozzetti}, {Fischer}, {Butler},
  {Marcy}, {Stefanik}, {Sasselov}, {L{\'a}z{\'a}r}, {Papp}, \&
  {S{\'a}ri}}]{2007ApJ...666L.121T}
{Torres}, G., {Bakos}, G.~{\'A}., {Kov{\'a}cs}, G., {et~al.} 2007, \apjl, 666,
  L121

\bibitem[{{Van Eylen} {et~al.}(2016){Van Eylen}, {Albrecht}, {Gandolfi}, {Dai},
  {Winn}, {Hirano}, {Narita}, {Bruntt}, {Prieto-Arranz}, {B{\'e}jar}, {Nowak},
  {Lund}, {Palle}, {Ribas}, {Sanchis-Ojeda}, {Yu}, {Arriagada}, {Butler},
  {Crane}, {Handberg}, {Deeg}, {Jessen-Hansen}, {Johnson}, {Nespral}, {Rogers},
  {Ryu}, {Shectman}, {Shrotriya}, {Slumstrup}, {Takeda}, {Teske}, {Thompson},
  {Vanderburg}, \& {Wittenmyer}}]{2016arXiv160509180V}
{Van Eylen}, V., {Albrecht}, S., {Gandolfi}, D., {et~al.} 2016, \aj, 152, 143

\bibitem[{{Vidal-Madjar} {et~al.}(2003){Vidal-Madjar}, {Lecavelier des Etangs},
  {D{\'e}sert}, {Ballester}, {Ferlet}, {H{\'e}brard}, \&
  {Mayor}}]{2003Natur.422..143V}
{Vidal-Madjar}, A., {Lecavelier des Etangs}, A., {D{\'e}sert}, J.-M., {et~al.}
  2003, \nat, 422, 143

\bibitem[{{Vidal-Madjar} {et~al.}(2004){Vidal-Madjar}, {D{\'e}sert},
  {Lecavelier des Etangs}, {H{\'e}brard}, {Ballester}, {Ehrenreich}, {Ferlet},
  {McConnell}, {Mayor}, \& {Parkinson}}]{2004ApJ...604L..69V}
{Vidal-Madjar}, A., {D{\'e}sert}, J.-M., {Lecavelier des Etangs}, A., {et~al.}
  2004, \apjl, 604, L69

\bibitem[{{Vogt} {et~al.}(1994){Vogt}, {Allen}, {Bigelow}, {Bresee}, {Brown},
  {Cantrall}, {Conrad}, {Couture}, {Delaney}, {Epps}, {Hilyard}, {Hilyard},
  {Horn}, {Jern}, {Kanto}, {Keane}, {Kibrick}, {Lewis}, {Osborne},
  {Pardeilhan}, {Pfister}, {Ricketts}, {Robinson}, {Stover}, {Tucker}, {Ward},
  \& {Wei}}]{1994SPIE.2198..362V}
{Vogt}, S.~S., {Allen}, S.~L., {Bigelow}, B.~C., {et~al.} 1994, in \procspie,
  Vol. 2198, Instrumentation in Astronomy VIII, ed. D.~L. {Crawford} \& E.~R.
  {Craine}, 362

\bibitem[{{Wittenmyer} {et~al.}(2011){Wittenmyer}, {Endl}, {Wang}, {Johnson},
  {Tinney}, \& {O'Toole}}]{Wittenmyer:2011}
{Wittenmyer}, R.~A., {Endl}, M., {Wang}, L., {et~al.} 2011, \apj, 743, 184

\bibitem[{{Zhou} {et~al.}(2016{\natexlab{a}}){Zhou}, {Latham}, {Bieryla},
  {Beatty}, {Buchhave}, {Esquerdo}, {Berlind}, \& {Calkins}}]{Zhou:2016}
{Zhou}, G., {Latham}, D.~W., {Bieryla}, A., {et~al.} 2016{\natexlab{a}},
  \mnras, 460, 3376

\bibitem[{{Zhou} {et~al.}(2016{\natexlab{b}}){Zhou}, {Rodriguez}, {Collins},
  {Beatty}, {Oberst}, {Heintz}, {Stassun}, {Latham}, {Kuhn}, {Bieryla}, {Lund},
  {Labadie-Bartz}, {Siverd}, {Stevens}, {Gaudi}, {Pepper}, {Buchhave},
  {Eastman}, {Col{\'o}n}, {Cargile}, {James}, {Gregorio}, {Reed}, {Jensen},
  {Cohen}, {McLeod}, {Tan}, {Zambelli}, {Bayliss}, {Bento}, {Esquerdo},
  {Berlind}, {Calkins}, {Blancato}, {Manner}, {Samulski}, {Stockdale},
  {Nelson}, {Stephens}, {Curtis}, {Kielkopf}, {Fulton}, {DePoy}, {Marshall},
  {Pogge}, {Gould}, {Trueblood}, \& {Trueblood}}]{2016arXiv160703512Z}
{Zhou}, G., {Rodriguez}, J.~E., {Collins}, K.~A., {et~al.} 2016{\natexlab{b}},
  \aj, 152, 136

\end{thebibliography}

\end{document}